\documentclass[journal=jpcafh,manuscript=article]{achemso}
\usepackage[utf8]{inputenc}

\usepackage{longtable}
\usepackage{multirow}
\usepackage{amsmath}
\usepackage{subfigure}
\usepackage[usenames,dvipsnames]{xcolor}
\usepackage{soul}
\usepackage{bm}

\usepackage{amssymb}
\usepackage{siunitx}
\usepackage{multicol} \usepackage{wrapfig} \usepackage{enumitem}
\usepackage{bm} \usepackage{outlines} 

\usepackage{url}
\usepackage{hyperref}

\usepackage{siunitx}\usepackage{booktabs}

\usepackage{xr}

\makeatletter
\newcommand*{\addFileDependency}[1]{
  \typeout{(#1)}
  \@addtofilelist{#1}
  \IfFileExists{#1}{}{\typeout{No file #1.}}
}
\makeatother

\newcommand*{\myexternaldocument}[1]{%
    \externaldocument{#1}%
    \addFileDependency{#1.tex}%
    \addFileDependency{#1.aux}%
}

\myexternaldocument{si}

\SectionNumbersOn

\author{Valerii Andreichev} \affiliation[University of
  Basel]{Department of Chemistry, University of Basel,
  Klingelbergstrasse 80, CH-4056 Basel, Switzerland.}

\author{Silvan K\"aser} \affiliation[University of Basel]{Department of
  Chemistry, University of Basel, Klingelbergstrasse 80, CH-4056
  Basel, Switzerland.}\altaffiliation{Present Address: Roche Pharma
  Research and Early Development, Pharmaceutical Sciences, Roche
  Innovation Center Basel, F. Hoffmann-La Roche Ltd, Basel,
  Switzerland}

\author{Erica L. Bocanegra} \affiliation[Yale University]{Sterling Chemistry
  Laboratory, Yale University, New Haven, Connecticut 06520, United
  States}

\author{Madeeha Salik} \affiliation[Yale University]{Sterling Chemistry
  Laboratory, Yale University, New Haven, Connecticut 06520, United
  States}

\author{Mark A. Johnson} \affiliation[Yale University]{Sterling Chemistry
  Laboratory, Yale University, New Haven, Connecticut 06520, United
  States}

\author{Markus Meuwly} \affiliation[University of Basel]{Department of
  Chemistry, University of Basel, Klingelbergstrasse 80, CH-4056
  Basel, Switzerland.}  \email{m.meuwly@unibas.ch}

\title{Dynamics of Protonated Oxalate from Machine-Learned Simulations
  and Experiment: Infrared Signatures, Proton Transfer Dynamics and
  Tunneling Splittings}

\begin{document}

\begin{abstract}
The infrared spectroscopy and proton transfer dynamics together with
the associated tunneling splittings for H/D-transfer in oxalate are
investigated using a machine learning-based potential energy surface
(PES) of CCSD(T) quality, calibrated against the results of new
spectroscopic measurements. Second order vibrational perturbation
calculations (VPT2) very successfully describe both the framework and
H-transfer modes compared with the experiments. In particular, a new
low-intensity signature at 1666 cm$^{-1}$ was correctly predicted from
the VPT2 calculations. An unstructured band centered at 2940 cm$^{-1}$
superimposed on a broad background extending from 2600 to 3200
cm$^{-1}$ is assigned to the H-transfer motion. The broad background
involves a multitude of combination bands but a major role is played
by the COH-bend. For the deuterated species, VPT2 and molecular
dynamics simulations provide equally convincing assignments, in
particular for the framework modes. Finally, based on the new PES the
tunneling splitting for H-transfer is predicted as $\Delta_{\rm H} =
35.0$ cm$^{-1}$ from ring polymer instanton calculations using
higher-order corrections. This provides an experimentally accessible
benchmark to validate the computations, in particular the quality of
the machine-learned PES.
\end{abstract}

\section{Introduction}
Vibrational spectroscopy is a nondestructive technique to characterize
the structure and dynamics of chemical and biological materials in the
gas- and condensed-phase.\cite{bakker:2010,gaigeot:2010,kraka:2020}
One of the main applications in chemistry is the identification and
classification of compounds which is essential in analytical,
materials, and synthetic chemistry. Of prime importance in the
practical application of vibrational spectroscopy to concrete problems
is the ability to assign spectral responses to the underlying motions
in the molecule or material of interest. This can sometimes be
achieved through analogy with related compounds for which assignments
of spectra to motions have already been made.\cite{MM.diglyme:2007}
Alternatively, isotopic substitutions can be used to assist with this
assignment which, however, can be tedious from a preparatory
perspective.\\

\noindent
One of the most direct ways to arrive at assignment and identification
of spectral patterns is through computation. Vibrational spectroscopy
is an area where ``experiment'' and ``theory/computation'' meet in a
most natural and direct fashion. However, fruitful interplay between
these two approaches requires high-level computer methods which
typically need to go beyond the harmonic approximation. It is
generally believed that calculations at the coupled cluster with
singles, doubles and perturbative triples (CCSD(T)) are needed for
reliable computational vibrational spectroscopic work.\cite{qu:2019}
Also, the nuclear dynamics needs to be described at a sufficiently
advanced level. Standard normal mode analysis is often sufficient to
obtain an overview of the spectral features. If quantitative agreement
with experiment is sought, more advanced techniques such as
vibrational perturbation theory (VPT2)\cite{barone:2010} or spectra
from explicit molecular dynamics (MD) simulations are needed,
though.\cite{gaigeot:2010,MM.review:2020}\\

\noindent
A chemically particularly interesting motif constitute hydrogen bonds
and shared hydrogen atoms/protons between an acceptor and a
donor. H-bonds and hydrogen/proton transfer (HT/PT) play important
roles in governing molecular structure, stability, and dynamics and
their energetics and dynamics is of fundamental importance in biology
and chemistry.\cite{cleland:1994,warshel:1995,herschlag:1996} HT/PT is
primarily influenced by the height of the barrier which is, however,
difficult to determine reliably through direct
experimentation. Possibilities include high resolution spectroscopy
where the splitting of spectral lines can provide information about
the barrier height,\cite{leutwyler:2002} or nuclear magnetic resonance
(NMR) experiments\cite{tobin:1994,limbach:1994} whereas kinetic
isotope effects or shifts of bands in the vibrational spectrum alone
can not be used directly to determine the energetics for PT. Proton
transfer in systems containing X$-$H$^*\cdots $Y motifs - where X and
Y are the donor and acceptor atoms, respectively, and H$^*$ is the
transferring hydrogen - can lead to characteristically broadened
features in vibrational spectra.\cite{bondesson:2007} This broadening
reflects strong coupling between the X$-$H stretch and other framework
modes of the environment and structural
heterogeneity.\cite{wolke:2015} The broadening can even persist down
to low temperatures and cooling the species does not necessarily lead
to sharper bands.\cite{johnson:2013,johnson:2014}\\

\noindent
In experiments, the broadening of spectral lines in X$-$H$^*\cdots $Y
species has been extensively observed and reported in both the liquid
and the gas phase\cite{bakker:2010}. For example, in liquid water, the
maximum in the OH-stretching spectrum shifts to the red as the
temperature decreases from $47^{\circ}$C to $-6^{\circ}$C, with
concomitant increase of the intensity by 16\% even without hydrogen or
proton transfer to take place.\cite{brubach:2005} In the high-density
liquid, the line shapes in the OH stretching spectra of supercritical
methanol was found to be sensitive to the hydrogen bonding network
which depends on temperature\cite{bakker:2010}.\\

\noindent
Spectral shifts and their broadening are even more pronounced in
systems where a transferring proton H$^*$ is shared by a donor and
acceptor moiety and potentially provides information about the proton
transfer energetics. An empirical relationship between the position of
the infrared (IR) absorption and the height of the proton transfer
barrier has been found in combined computational/experimental
investigations of acetylacetone\cite{MM.acac:2015} and formic acid
dimer\cite{MM.fa:2016}. Earlier studies of the infrared spectroscopy
of protonated ammonia dimer have also established that the
IR-signatures are broad and correlated with the barrier height for
proton transfer\cite{MM.amm:2002}.  \\

\begin{figure}[h!]
\centering
\includegraphics[width=0.5\textwidth]{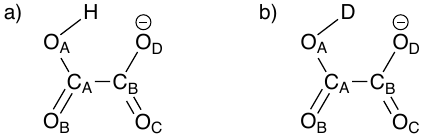}
\caption{Atom labels for OxH and OxD molecules used for
  internal coordinates in power spectra.}
\label{fig:labels}
\end{figure}

\noindent
Here, a high-level theoretical study is combined with improved
spectroscopic measurements to elucidate the vibrational dynamics at
play in the protonated oxalate (HO$_2$CCO$_2^-$) anion, hereafter
referred to as OxH and OxD for the hydrogenated and deuterated forms
of the protonated dianion, see Figure~\ref{fig:labels}. OxH is a
compelling system to study the dynamics and ensuing spectroscopic
signatures of a shared proton coupled to framework modes of the
surrounding molecular scaffold.\cite{wolke:2015,MM.oxa:2017} Previous
computational work concerned the kinetic isotope effects from
electronic structure calculations,\cite{truong:1991} the solution
dynamics of hydrated $p-$Oxa,\cite{kabelac:2016} and the IR
spectroscopy in the gas phase using a reactive empirical energy
function.\cite{MM.oxa:2017} The experimental IR spectrum features a
rather diffuse/broad signal between 2600 and 2900~cm$^{-1}$. One
potential source is strong mechanical coupling between the
high-frequency O-H stretch and (potentially) the low-frequency COH
bends and/or skeletal deformations.\\

\noindent
Studying the dynamics and spectroscopy of a transferring proton from
MD simulations requires energy functions capable of describing
bond-formation and bond-breaking. One such method is molecular
mechanics with proton transfer (MMPT) which combines an accurate
representation of the energetics for the transferring proton and a
lower-level empirical force field for the remaining degrees of freedom
which allows long simulation
times.\cite{MM.mmpt:2008,MM.mmpt:2011,MM.ma:2010} Using ``PES
morphing''\cite{MM.morph:1999} provides flexibility to adapt
characteristics such as the barrier height and position of the
minima. More recently, alternative ways to represent high-dimensional
potential energy surfaces have been explored. One of them is based on
neural networks and was applied to bulk silicon or the H+HBr
reaction.\cite{behler:2007,komanduri:2009} The evaluation of the
neural network, once trained, is orders of magnitude faster than the
underlying reference quantum chemical calculations the model was
trained on. To reach the highest levels of quantum chemical theory,
such as coupled cluster-level, which is required for spectroscopic
studies, Transfer learning (TL) can be used. Such approaches have been
shown to be data efficient for obtaining validated high quality
PESs\cite{dral2020hierarchical,mm.anharmonic:2021,kaser2022transfer,MM.tlma:2022,zaverkin2023transfer,chen2023data,nandi2023ring,kaser2023microsecond,nandi2024delta,MM.tl:2025}. TL
is based on the notion that the topography of the energy landscape of
a molecular system is very similar even at different levels of
theory. This allows fine-tuning the learnable parameters of a neural
network trained on a lower level of theory using small amounts of data
calculated at a high level. Here, the MP2 level of theory provides a
computationally efficient route to a broad coverage of the energy
landscape, while the smaller high-accuracy dataset allows more precise
modeling.\\

\noindent
With such ``vetted'' PESs, it should be even possible to {\it predict}
experimentally accessible observables. Similar to the shared proton
situation in malonaldehyde or tropolone, OxH should exhibit tunneling
splitting for the (quantum) motion between the two symmetrical
wells. Quantum mechanical tunnelling is a key concept in chemistry and
physics and is well established,\cite{gol2021tunneling} and includes,
{\it inter alia}, electron transfer in
solution\cite{chandler1998electron,menzeleev2011direct,fang2020revisiting,lawrence2020confirming},
enzymatic
reactions\cite{hammes2006hydrogen,hu2014extremely,rommel2012role}, and
the structure and dynamics of
water\cite{keutsch2001water,richardson2016concerted,cvitavs2021quantum,ceriotti2016nuclear}. In
contrast to the often challenging measurement of chemical rates with
high precision, tunnelling splittings can be accurately measured
through spectroscopic methods, e.g., using microwave
\cite{rowe1976intramolecular} or infrared spectroscopy
\cite{ortlieb2007proton}. Tunneling splittings provide valuable
benchmarks for validating state-of-the-art theoretical methodologies,
which require both an accurate PES and an accurate treatment of the
resulting nuclear Hamiltonian. \cite{cruzan:1996} The calculation of
accurate tunneling splittings for multidimensional systems using fully
quantum mechanical methods is a challenging task. The ring-polymer
instanton (RPI) approach is a semiclassical approximation method for
tunnelling calculations which is capable of scaling well with system
size.
\cite{richardson2011rpi,richardson2018review,richardson2016concerted}
However, instanton theory makes a local harmonic approximation around
the optimal tunnelling pathway, resulting in deviations from
quantum-mechanical benchmarks that can reach 20\%
\cite{richardson2018perspective}. This error can be mitigated by
perturbatively correcting the results using information about the
third and fourth derivatives along the path
\cite{lawrence2023perturbatively}. It has been demonstrated that
correction of a low level potential energy surface to a higher level
using TL, in combination with ring-polymer instanton, can be used to
quantitatively characterise tunnelling splittings in such systems as
malonaldehyde \cite{MM.tlma:2022} and larger systems consisting of
15-atom tropolone and 12-atom propiolic acid-formic acid dimer (PFD)
\cite{MM.tl:2025}.  \\

\noindent
The present work is structured as follows. First, the computational
and experimental methods are described, followed by a validation of
the machine learned PESs. Then, results on the IR spectroscopy of OxH
and OxD, the H-transfer dynamics and HT tunneling splittings are
presented. Finally, conclusions are drawn.\\

\section{Methods}

\subsection{Machine Learning-Based PESs}
Machine learning is used to learn a full-dimensional, reactive
potential energy surface (PES) for OxH based on \textit{ab initio}
reference data of MP2/aug-cc-pVTZ quality. The data, available from
previous work\cite{kaeser2025kernn}, was used to train a
message-passing neural network
(PhysNet)\cite{MM.physnet:2019}. Structures for OxH were sampled by
running $NVT$ MD simulations at multiple temperatures ([100, 300, 500,
  1000, 1500]~K) using the semiempirical GFN2-xTB
method\cite{bannwarth2019gfn2} (2500 structures each except for 1500~K
for which only 1000 were used). The region around the proton transfer
transition state was sampled with a constraining harmonic potential
(1500 structures) and at $T = 500$~K. Additionally, normal mode
sampling\cite{smith2017ani} at progressively higher temperatures
([100, 300, 500, 1000, 1500, 2000]~K) was carried out for both the
optimized and transition state structure of OxH (800 for each
$T$). This yields a total of 22100 structures, for which energies,
forces and dipole moments were determined at the MP2/aug-cc-pVTZ level
of theory using MOLPRO\cite{MOLPRO} and served as the initial
reference data set. PhysNet was then trained following the procedures
described in detail in Reference~\citenum{MM.physnet:2019} using the
same hyperparameter values and a 80/10/10~\% split of the data as
training/validation/test sets. Two independent PhysNet models were
trained, which were subsequently used for active
learning\cite{unke2021machine}.\\

\noindent
Adaptive sampling\cite{csanyi2004learn} and diffusion Monte Carlo
simulations (DMC)\cite{kosztin1996introduction} were run to ensure the
robustness of the PES and to locate so-called ``holes''. Adaptive
sampling was done at 1000~K in the $NVT$ ensemble and structures (to
be included in training the next models) were detected if the
predictions of the two independent models differed by more than
0.5~kcal/mol. All DMC simulations employed a GPU implementation of the
DMC algorithm\cite{kaser2022transfer} for PhysNet and were run with
300 walkers, a step size of 5 a.u. and for a total of 60000
steps. Unphysical structures were saved if the predicted energy was
lower than the energy of the minimum energy structure. During the
first active learning cycle, adaptive sampling provided 80 structures
and 20 defective structures were located from DMC
simulations. \textit{Ab initio} MP2 data for the 100 structures was
determined and added to the MP2 data set before retraining PhysNet. A
second active learning cycle was performed, however, no additional
deficient structures were found. The final and robust MP2-based PES
constitutes the ``base model''.\\

\noindent
To improve the performance of the base model for molecular
spectroscopy, TL was used to refine and elevate the PES from the
MP2/aug-cc-pVTZ to the gold-standard CCSD(T)/aug-cc-pVTZ level of
theory.\cite{taylor2009transfer,pan2009survey,smith2019approaching}
The TL data set contains a total of 2688 structures selected
semi-randomly from the original data set. It comprises 1067 structures
that were sampled using MD simulations at 1000~K, 960 and 661
structures obtained from normal mode sampling around the minimum and
transition state, respectively, for which energies, forces and dipole
moments were determined at the CCSD(T)/aug-cc-pVTZ level of theory
using MOLPRO.\cite{MOLPRO} These structures and corresponding quantum
chemical information were then used to fine-tune the learnable
parameters of the MP2 PES with the same hyperparameters as for the
low-level training except for the learning rate and the weighting
hyperparameters in the loss function. The learning rate was reduced to
$10^{-4}$ and the force, dipole and total charge weights were all set
to one.\\

\subsection{Molecular Dynamics Simulations}
The molecular dynamics simulations were run using the atomic
simulation environment (ASE)\cite{larsen2017atomic} with energies and
forces obtained from PhysNet. The structure of OxH was first
optimized. Next, random momenta were drawn from a Maxwell-Boltzmann
distribution corresponding to $T = [300, 600, 900]$~K, which were
assigned to the atoms. The MD simulations were carried out in the
\textit{NVE} ensemble using the Velocity Verlet
algorithm\cite{verlet:1967} and a time step of $\Delta t = 0.2$~fs to
conserve energy as bonds involving hydrogen were flexible. The
molecular system was equilibrated for 2~ps, followed by production
simulations of 200~ps. For each temperature and system (OxH and OxD),
500 independent trajectories were run using different initial
momenta. This accumulated to a total of 100~ns simulation time per
temperature $T$.\\

\subsection{Infrared Spectroscopy}
IR spectra were calculated from the Fourier transform of the
dipole-dipole auto-correlation function\cite{gordon:1968,berne:2000}
according to
\begin{equation}
  I(\omega) n(\omega) \propto Q(\omega) \cdot \mathrm{Im}\int_0^\infty
  dt\, e^{i\omega t} 
  \sum_{i=x,y,z} \left \langle \boldsymbol{\mu}_{i}(t)
  \cdot {\boldsymbol{\mu}_{i}}(0) \right \rangle.
\label{eq:IR}
\end{equation}
To account for quantum effects, the Fourier transform was further
adjusted by applying a quantum correction factor\cite{lawrence:2005}
\[
Q(\omega) = \tanh\left(\frac{\beta \hbar \omega}{2}\right).
\]
\\

\subsection{Ring Polymer Instanton Calculations}
The RPI method is a semiclassical approximation for computing
tunneling splittings in molecular systems.\cite{tunnel,InstReview} RPI
locates the optimal tunneling pathway, known as the instanton, and is
defined as an imaginary-time $\tau\rightarrow\infty$ path connecting
two degenerate wells which minimizes the total action, $S$. The path
is constructed by optimizing a path discretized into $N$ ring-polymer
beads and taking the limit $N\rightarrow\infty$ (typically $N \sim
1000$ is sufficient for convergence). The potential $U_N$ of a ring
polymer is given by
\begin{align}
  \label{eq:rpi_polymer_potential}
    U_N(\bm{x};\beta) = \sum_{i=1}^N V(\mathbf{x}_i) +
    \frac{1}{2(\beta_N\hbar)^2}\sum_{i=1}^N |\mathbf{x}_{i+1} -
    \mathbf{x}_i|^2 \equiv \frac{S({\bm x})}{\beta_N \hbar}
\end{align}
with $\bm{x} = \left( \mathbf{x}_1, \dots, \mathbf{x}_N\right)$ being the
mass-scaled coordinates of the beads, $\beta = \frac{1}{k_b T}$ and
$\beta_N = \beta/N$. The first term in
Equation~\ref{eq:rpi_polymer_potential} corresponds to the sum over
all single bead potentials and the second term represents the harmonic
springs with frequency $1/(\beta_N\hbar)$ that connect adjacent
beads. As the potential $U_N$ and the action $S$ are related by
division with $(\beta_N\hbar)$, the action $S$ contains information
\emph{along} the instanton path (IP). A detailed description of the
method is provided, \textit{e.g.}, in References~\citenum{tunnel} and
\citenum{InstReview}.\\

\noindent
Within standard RPI (sRPI) theory, fluctuations around the path are
computed to second order and the information is combined into the term
$\Phi$, \textit{i.e.}, this is based on information \emph{around} the
IP. This requires Hessians at each of the beads. The leading-order
tunneling splitting in a double-well system is
\begin{align}
    \Delta_{\rm RPI} = \frac{2\hbar}{\Phi} \sqrt\frac{S}{2\pi\hbar} \,
    \mathrm{e}^{-S/\hbar}.
\label{sieq:rpi_splitting} 
\end{align}
For malonaldehyde and formic acid dimer sRPI tunneling splittings on
high quality PESs were within $\sim 20$ \% of the experimental
values\cite{richardson:2017,MM.tlma:2022,MM.tl:2025}. This error
arises due to approximations underlying sRPI
theory\cite{richardson2018perspective} and remaining deficiencies in
the PES because $\Delta_{\rm RPI}$ depends exponentially on the action
$S$, see Equation \ref{sieq:rpi_splitting}.  One limitation of the
sRPI method for determining tunneling splittings is that fluctuations
around the instanton are treated harmonically, \textit{i.e.}, the
Hessians enter via $\Phi$.\cite{InstReview} To a good approximation,
it is expected that this captures the dominant tunneling contribution,
except for cases in which anharmonic effects perpendicular to the
instanton are significant or where the barrier is low. For this
reason, a perturbatively corrected RPI (pcRPI) theory was recently
developed\cite{lawrence2023perturbatively}, that accounts for
anharmonicity by including information from the third and fourth order
derivatives of the potential along the instanton. The tunneling
splitting obtained from pcRPI theory is denoted as $\Delta_{\rm PC}$
in the following, which, in practice is obtained from scaling
$\Delta_{\rm RPI}$ with a correction factor $c_{\rm PC}$ according to
$\Delta_{\rm PC} = c_{\rm PC}\cdot \Delta_{\rm RPI}$.\\

\subsection{Experimental}
The vibrational predissociation spectrum of oxalate with two
H$_2$-tags was recorded on the Yale triple focusing photofragmentation
time-of-flight (TOF) mass spectrometer that has been described
previously.\cite{farrar:1988, menges:2019, yang:2018} 1 mM solutions
of oxalic acid in pure acetonitrile were prepared and the ionic
species were generated via electrospray ionization (ESI) of the oxalic
acid solutions in negative ion mode. ESI results in deprotonation of
one of the carboxylic acid head groups, generating the oxalate
anion. Ions are then transferred via radiofrequency guides to a
temperature-controlled (5 - 40 K), 3D (Paul) ion trap. A few ms before
the ions arrive in the trap, He buffer gas doped with ~20\% H$_2$ was
pulsed into the trap where collisions with He thermalize the ions to
the trap temperature, allowing H$_2$ tags to condense onto the
ions. \cite{wolk:2014, mikosch:2004, boyarkin:2014, wester:2009,
  asvany:2009} After $\sim 90$ ms, the tagged ions are ejected from
the trap into the TOF region of the instrument. At a transient focus
of the TOF, the ions are excited with an IR pulse generated from a
LaserVision OPO/OPA (5 ns, 2 - 20 mJ/pulse) in the 800 – 4200
cm$^{-1}$ range. Photofragments were generated when the ions absorb a
single resonant IR photon, evaporating the H$_2$ tags. The
photofragment yield was measured as a function of photon energy to
generate an infrared photodissociation (IRPD) spectrum in a linear
action mode. The reported spectra are normalized to the laser fluence
to account for the wavelength dependence of the output power.\\

\section{Results}

\subsection{Validation of the Potential Energy Surfaces}
{\bf The MP2 PES:} The out-of-sample errors of the MP2 base model are
reported in Table~\ref{tab:outofsample_errors_tl} and their
correlation and signed errors are shown in
Figure~\ref{sifig:errcorr}. The energy barrier for hydrogen transfer
on the PhysNet PES is 2.355~kcal/mol and differs from the \textit{ab
  initio} MP2/aug-cc-pVTZ barrier by only $\Delta E_a =
8\cdot10^{-4}$~kcal/mol. Harmonic frequencies, obtained from
diagonalizing the Hessian matrix characterize the shape of the PES
around stationary points. PhysNet achieves a MAE($\omega$) for the
harmonic frequencies at the global minimum and the TS for H-transfer
of 0.09 and 0.11~cm$^{-1}$ with respect to the \textit{ab initio}
harmonic frequencies. All harmonic frequencies are given in
Table~\ref{sitab:harmfreq}.\\

\noindent
{\bf The Transfer-Learned CCSD(T) PES:} Transfer-learning to the
CCSD(T) level of theory used only 10 \% of the data underlying the
MP2-base model. The out-of-sample errors for the TL-PES are given in
Table~\ref{tab:outofsample_errors_tl}. Their correlation and signed
errors are shown in Figure~\ref{sifig:errcorr_ccsd}. The energy
barrier for hydrogen transfer on the PhysNet PES is 3.351~kcal/mol and
differs from the \textit{ab initio} CCSD(T)/aug-cc-pVTZ barrier by
$\Delta E_a = 7\cdot10^{-4}$~kcal/mol. The MAE($\omega$) for the
minimum and H-transfer TS between harmonic frequencies from the TL-PES
and normal modes from explicit CCSD(T)/aVTZ calculations are 0.34 and
0.57~cm$^{-1}$ with respect to the \textit{ab initio} harmonic
frequencies. All harmonic frequencies are given in
Table~\ref{sitab:harmfreq_tl}.\\

\begin{table}[h]
\begin{tabular}{lll}\toprule
\textbf{kcal/mol(/\AA)}      & \textbf{PhysNet (MP2)} & \textbf{PhysNet (CCSD(T))}  \\\midrule
MAE($E$) & 0.009 & 0.005\\
RMSE($E$) & 0.047 & 0.012\\
MAE($F$) & 0.065 & 0.023 \\
RMSE($F$) & 0.459 & 0.065 \\
$1-R^2$ & 5.9E-06 & 7.8E-07 \\
$E_a$ & 2.355 & 3.351 \\
$\Delta E_a$ & 8E-04 & 7E-04 \\
MAE($\omega$) (cm$^{-1}$) & 0.09 & 0.34 \\
\bottomrule
\end{tabular}
\caption{Mean absolute (MAE) and root mean squared errors (RMSE) for
  energies $E$ and forces $F$ on a hold-out test set containing 2220
  and 270 random OxH structures for MP2 and CCSD(T) level of theory,
  respectively. In addition, the deviation $1-R^2$ of the correlation
  coefficient from 1, the barrier height $E_a$ of the PhysNet model
  and the deviation $\Delta E_a$ from the corresponding electronic
  structure calculation is reported. MAE$(\omega)$ is the difference
  between the normal mode frequencies calculated from the ML-PES and
  at the respective level of theory. For normal modes see also Tables
  \ref{sitab:harmfreq} and \ref{sitab:harmfreq_tl}.}
\label{tab:outofsample_errors_tl}
\end{table}

\subsection{Infrared Spectroscopy}
\begin{figure}[h!]
\centering
\includegraphics[width=1.0\textwidth]{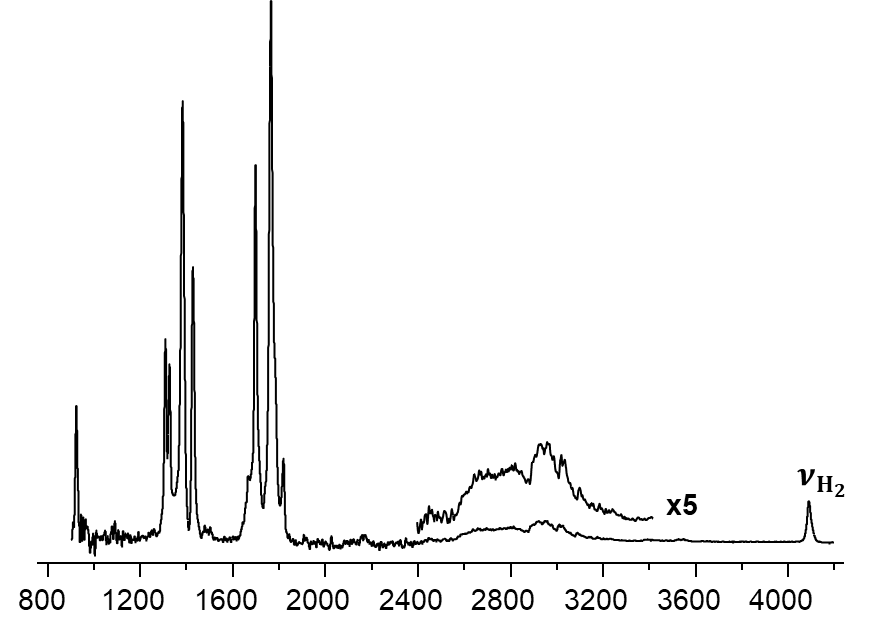}
\caption{Experimental predissociation vibrational spectrum of
  doubly-H$_2$ tagged oxalate. For clarity, the spectrum in the range
  2400 - 3400 cm$^{-1}$ is expanded by a factor of 5 in the inset. The
  sharper feature at 4090 cm$^{-1}$ is due to the nominally IR
  forbidden H$_2$ fundamental, which appears --72 cm$^{-1}$ below the
  vibrational quantum in isolated (bare) H$_2$.\cite{moss:2024}}
\label{fig:exp_oxh}
\end{figure}

The infrared spectrum of OxH-(D$_2$)$_2$ recorded for this study,
hereafter denoted IR2025, is displayed in Figure
\ref{fig:exp_oxh}. This spectrum is characterized by a number of
sharper bands for the region between 1200 and 2000 cm$^{-1}$ and a
considerably broader feature between 2500 and 3200 cm$^{-1}$. This
feature can be viewed as a band centered around 2950 cm$^{-1}$,
superimposed on a diffuse background absorption spanning over 700
cm$^{-1}$. Because there is only one hydrogen atom in OxH, the band in
the high-frequency region must be associated with fundamentals and
combination modes involving primarily the OH-stretching and
COH-bending modes.\\

\begin{figure}[h!]
\centering
\includegraphics[width=1.0\textwidth]{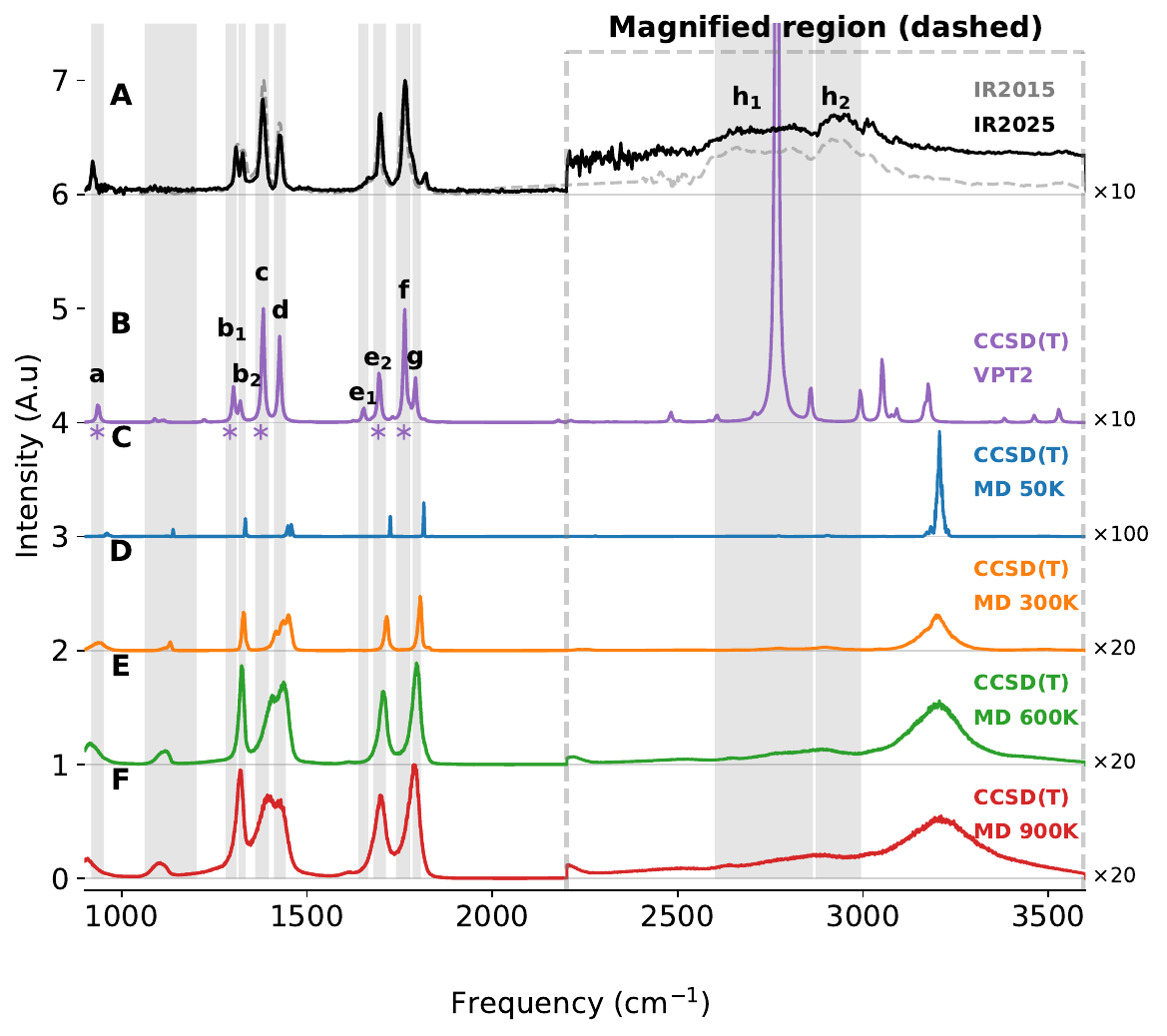}
\caption{Infrared spectra of OxH. Panel A: measured (H$_2$-tagged) gas
  phase spectra IR2015\cite{wolke:2015} (dashed grey) and IR2025 from
  the present work (solid black). Panel B: Spectrum from second-order
  vibrational perturbation theory, VPT2,\cite{barone:2010} using
  Lorentzian Broadening with ${\rm HWHM} = 4$ cm$^{-1}$ and bands
  labelled from {\bf a} to {\bf g} for identification. Asterisks
  indicate fundamental vibrations, see also Table
  \ref{sitab:oxh_assign}. Panels C to F: IR spectra at different
  simulation temperatures as indicated from the Fourier transform of
  the dipole-dipole autocorrelation function. Grey shaded areas
  highlight the position and width of the measured framework modes for
  direct comparison between computations and experiment.}
\label{fig:ir-sim}
\end{figure}

\noindent
One specific reason to revisit the previously published IR
spectrum\cite{wolke:2015} was to reassess the broad background between
2500 and 3200 cm$^{-1}$ and establish the relative intensities of the
bands across the entire spectrum. The complete spectrum also includes
the nominally forbidden fundamental of the H$_2$ stretch, (observed at
4090 cm$^{-1}$, labeled $\nu_{\rm H_2}$) which is red-shifted by --72
cm$^{-1}$ when attached to the molecular anion. The IR2015 (grey
dashed) and IR2025 (black solid) spectra are reported in Figure
\ref{fig:ir-sim}A. The plateau-like shape between 2550 to 2820
cm$^{-1}$ in the IR2015 spectrum features a more gradual increase in
IR2025 but overall, the two spectra are remarkably similar, see
Figure~\ref{sifig:zoomed_figure3}. The spectrum from the present work,
however, reveals a better defined shoulder near 1660 cm$^{-1}$ on the
red side of the strong band at 1696 cm$^{-1}$ than was evident in the
IR2015 spectrum (Figure~\ref{sifig:zoomed_figure3}). All other
features agree between the two spectra. One reason for the differences
between IR2015 and IR2025 concerned normalization with respect to
laser power which was much improved in IR2025.\\

\noindent
Second order vibrational perturbation theory\cite{barone:2010}
provides a useful way to go beyond the harmonic approximation
underlying normal mode analysis, also for H-bonded
systems.\cite{tschumper:2014} The VPT2 spectrum from using the TL-PES
is reported in Figure \ref{fig:ir-sim}B. Focusing first on the
framework modes below 2000 cm$^{-1}$ it is found that all features in
the measured spectrum line up well with the VPT2 spectrum (bands {\bf
  a} to {\bf g}; fundamentals {\bf a, b$_1$, c, e$_2$, f} indicated by
asterisks). This includes the multiplett-structures around 1360
cm$^{-1}$ (bands {\bf b$_1$}, {\bf b$_2$}, {\bf c}, and {\bf d}) and
centered at 1750 cm$^{-1}$ (bands {\bf e$_1$}, {\bf e$_2$}, {\bf f},
and {\bf g}). Most notably, a low-intensity signature {\bf e$_1$} at
1666 cm$^{-1}$ which was not clearly present in the IR2015 spectrum,
is manifest in the VPT2 calculations and the IR2025 spectrum, see
Figure \ref{sifig:zoomed_figure3}, and assigned to the $\nu_5 +
\nu_{10}$ combination band, see Table
~\ref{sitab:oxh_assign}. Furthermore, the low-frequency $\nu_9$ mode
(band {\bf a}) at 928 cm$^{-1}$ in the IR2025 spectrum is at 937
cm$^{-1}$ in the VPT2 calculations and a small feature at 1091
cm$^{-1}$ from the measurements and at 1090 cm$^{-1}$ from the VPT2
calculations (see entry with asterisk in Table \ref{sitab:oxh_assign},
not labelled). The VPT2 spectrum using the MP2 ML-PES is only
qualitatively correct, see Figure \ref{sifig:ir_hoxa_physnet}, and is
not suitable for quantitative analysis.\\

\noindent
The high-frequency part of the measured spectrum shows a wide
absorption ({\bf h$_1$}) between 2500 and 3200 cm$^{-1}$ onto which a
less broad signature {\bf h$_2$} centered at $\sim 2900$ cm$^{-1}$ is
superimposed. VPT2 calculations feature a considerable number of
discrete peaks covering the frequency range consistent with the
experiments. As the experiment does not probe a single conformation
but rather a distribution of structures, it is of interest to obtain
VPT2 frequencies for slightly perturbed minimum energy structures. For
this, the optimized oxalate structure was distorted by perturbing bond
lengths with a standard deviation of 0.01 \AA\/ followed by a VPT2
calculation for each structure. The Boltzmann-averaged spectrum, shown
in Figure \ref{sifig:oxoh_vpt2_averaged}, demonstrates that the
computed signal is more representative of the measured frequency
distribution than the VPT2 calculation for a single minimum energy
structure. It should be noted that VPT2 calculations use the harmonic
approximation as its zeroth-order approximation to which anharmonic
corrections are added. Hence, for strongly anharmonic modes or modes
involving Fermi resonances a VPT2 treatment may be
unreliable.\cite{skinner:2018,mm.anharmonic:2021}\\

\noindent
H-transfer during the dynamics will also be reflected in the infrared
spectroscopy. To include finite-temperature effects, MD simulations
using the TL-PES were carried out for temperatures between 50 K and
900 K, see Figures \ref{fig:ir-sim}C to F. The OH-stretch is found at
3200 cm$^{-1}$ and broadens considerably as temperature increases. The
blue shift away from the maximum of H-stretch mode at 2940 cm$^{-1}$
from experiments is well-understood and occurs because MD simulations
only sample a narrow range around the bottom of the well without
accessing the anharmonic parts of the PES, in particular for the
OH-stretch mode.\cite{suhm:2020}. As temperature increases, broader
signatures emerge which overlap with the peak positions from the VPT2
calculations and the measured spectrum.\\

\begin{figure}[h!]
\centering
\includegraphics[width=1.0\textwidth]{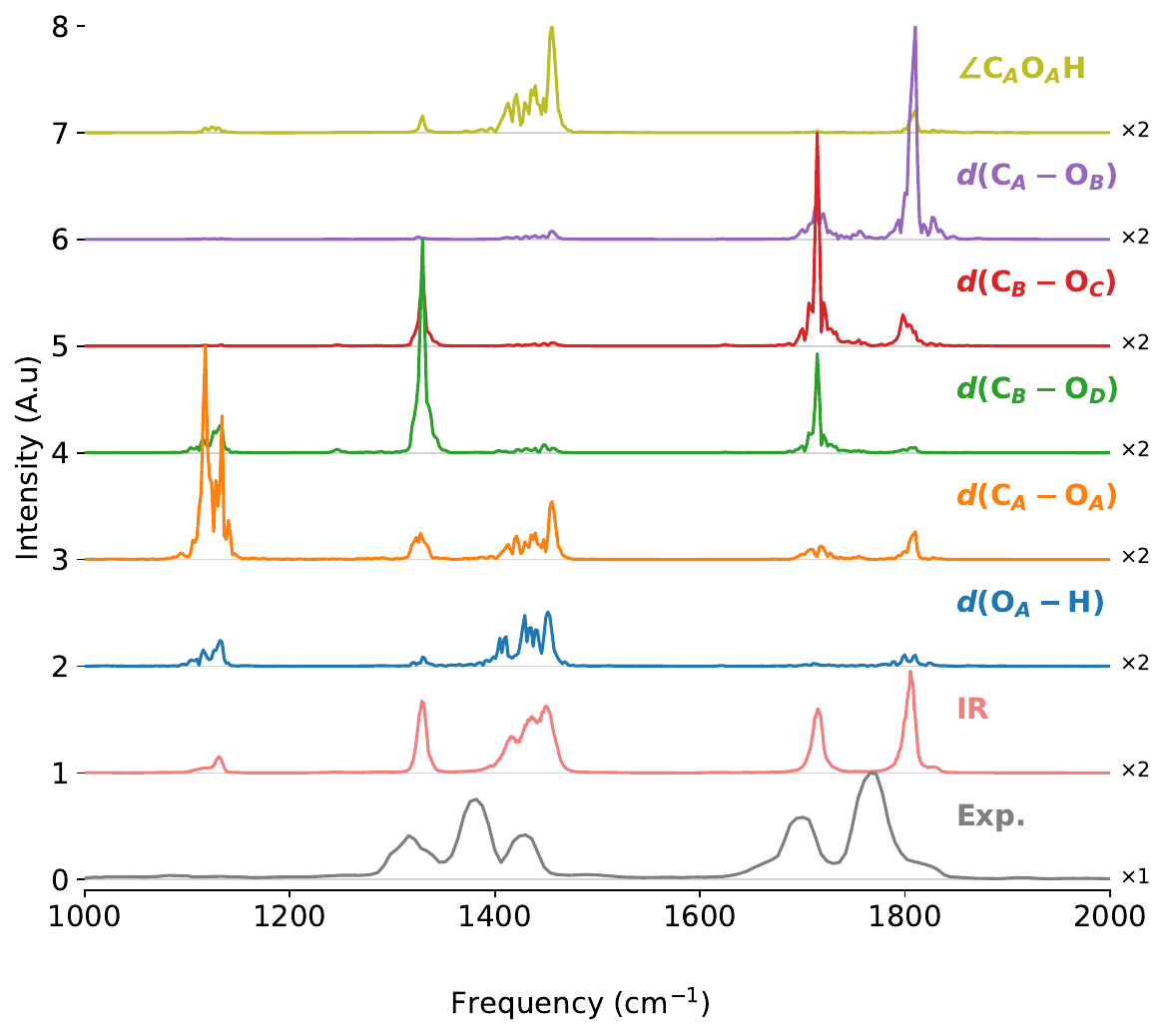}
\caption{Power spectra for OxH molecule at 300K for different internal
  coordinates (low-frequency range).}
\label{fig:ps300}
\end{figure}

\noindent
For partial assignment of the measured peaks, power spectra for
different internal coordinates were determined, see Figure
\ref{fig:ps300}. Due to coupling between different degrees of freedom,
straightforward and unambiguous assignments are not particularly
meaningful. Two infrared signatures at 1700 and 1770 cm$^{-1}$ are
clearly assigned in the IR spectrum and involve the C$_{\rm
  A}$-O$_{\rm A}$, C$_{\rm A}$-O$_{\rm B}$, C$_{\rm B}$-O$_{\rm C}$,
and C$_{\rm B}$-O$_{\rm D}$ distances. The peak to the red, at 1666
cm$^{-1}$ and part of the high-energy triplet, does not appear in the
IR and power spectra, but is clearly visible in the VPT2 calculations
which also report combination modes and hot bands, see Table
\ref{sitab:oxh_assign}. The triplet centered around 1360 cm$^{-1}$ is
also reproduced by the computed IR spectra and involves the C$_{\rm
  A}$O$_{\rm A}$H angle and all four CO-stretches.\\

\begin{figure}[h!]
\centering
\includegraphics[width=1.0\textwidth]{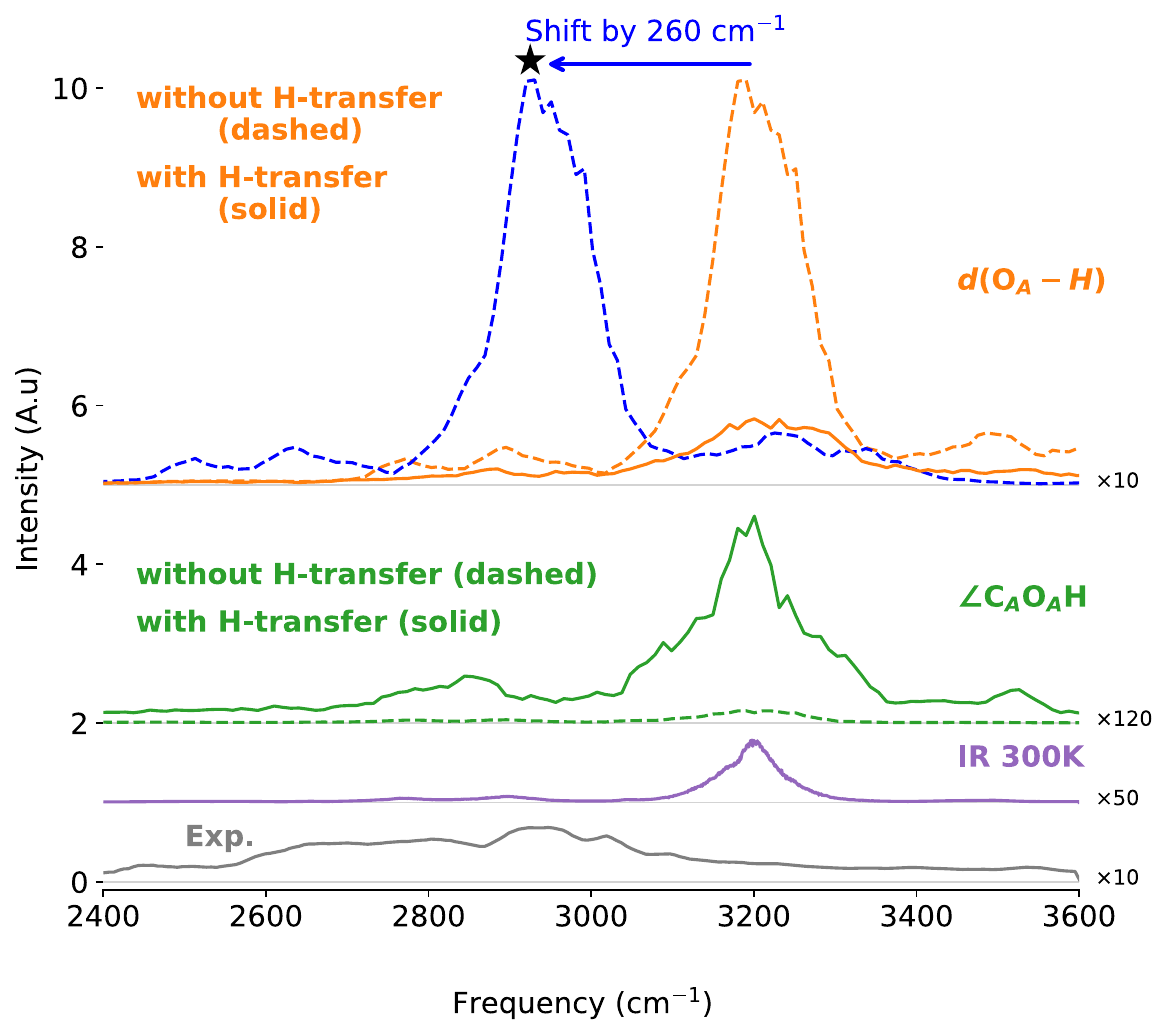}
\caption{Power spectra for the COH-bend (green) and OH-stretch
  (orange) from simulations at 600 K that with (solid line, 362
  trajectories) and without (dashed line, 68 trajectories) H-transfer
  during the dynamics. The indigo trace is the IR spectrum (300 K; no
  H-transfer). The blue dashed line is the shifted OH-stretch power
  spectrum to best overlap with the feature in the measured
  spectrum. For spectra at higher $T$ that feature H-transfer, see
  Figure \ref{fig:ir-sim}}
\label{fig:ps-oxh}
\end{figure}

\noindent
In order to further investigate the nature of the broad peak in the
2600-3200 cm$^{-1}$ region, MD trajectories run at 600 K were
analyzed. For this, separate IR and power spectra were generated for
trajectories that did or did not feature H-transfer, see Figures
\ref{fig:ps-oxh}, \ref{sifig:ps_600K_OxOH_low} and
\ref{sifig:ps_600K_OxOH_high}. The internal coordinates considered
were the C$_{\rm A}$-O$_{\rm A}$-H angle and the C$_{\rm A}$-O$_{\rm
  B}$, C$_{\rm B}$-O$_{\rm C}$, C$_{\rm B}$-O$_{\rm D}$, C$_{\rm
  A}$-O$_{\rm A}$, and O$_{\rm A}$-H separations. As already
mentioned, the blue shift of the OH-stretching vibration is
well-understood.\cite{suhm:2020,MM.fad:2022} However, the broad
[2600,3200] cm$^{-1}$ background from the measurements is rather
convincingly associated with signatures in the C$_{\rm A}$-O$_{\rm
  A}$-H angle especially for trajectories featuring
H-transfer. Interestingly, shifting the OH-power spectrum to the red
by --260 cm$^{-1}$ not only aligns the experimentally observed peak at
2940 cm$^{-1}$ but also the faint measured feature around 2450
cm$^{-1}$. For formic acid monomer,\cite{MM.tl:2022} the shift
required to align the computed with the measured infrared signature
arising from the OH-stretch was --290 cm$^{-1}$, which is consistent
with the observation for the OH-stretch in oxalate.\\

\noindent
To further corroborate this assignment, simulations were carried out
whereby the mass of the hydrogen atom ranged from $1 m_{\rm H}$ to $2
m_{\rm H}$ (i.e. $m_{\rm D}$) in increments of 0.2 mass units, see
Figure \ref{sifig:masses}. As expected, the spectral response shifts
to the red as the mass of the hydrogen atom increases. It is
interesting to note that with increasing temperature in Figure
\ref{fig:ir-sim} the region above 2600 cm$^{-1}$ gains intensity (note
also the different $y-$scaling factors). Analysis of the MD
trajectories shows that with increasing temperature the probability
for H-transfer between the oxygen atoms O$_{\rm A}$ and O$_{\rm D}$
increases. Hence, the power spectra for different internal coordinates
were separately analyzed for trajectories that did (dashed lines) and
did not (solid) feature hydrogen transfer, see Figure
\ref{fig:ps-oxh}. In the region of interest (above 2600 cm$^{-1}$) the
power spectrum of the COH bending motion (green trace) features clear
signals which are totally absent from simulations that do not feature
H-transfer. The OH-stretch motion (orange) is also clearly visible and
coupled to the COH-bend. \\

\begin{figure}[h!]
\centering \includegraphics[width=1.0\textwidth]{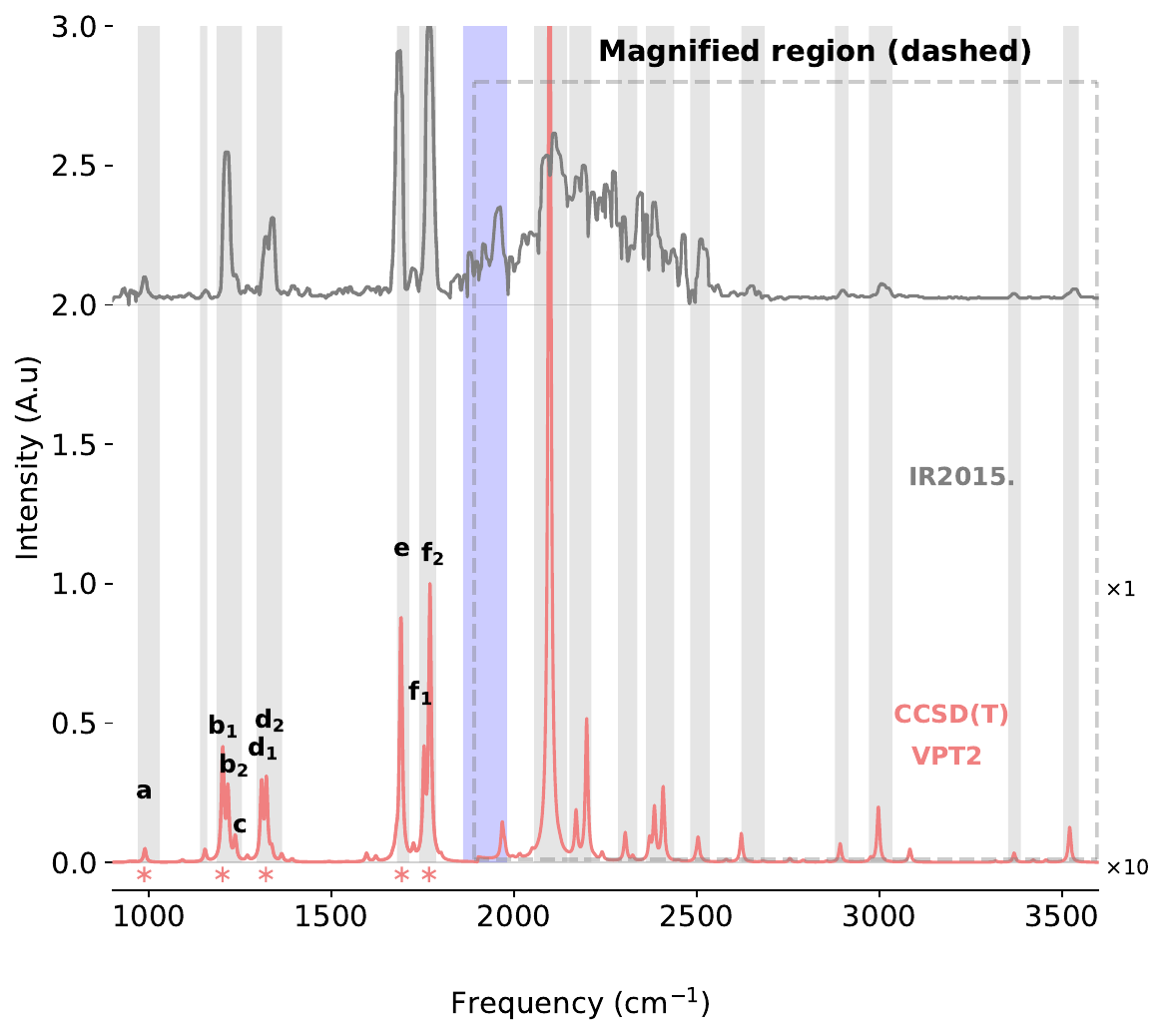}
\caption{Infrared spectra for OxD. The top trace reports the
  experimentally determined (H$_2$-tagged) gas phase
  spectrum\cite{wolke:2015} and the bottom trace (light coral) is the
  computed spectrum from second-order vibrational perturbation
  theory\cite{barone:2010} using the CCSD(T) ML-PES.}
\label{fig:oxod_vpt2}
\end{figure}

\subsection{Spectroscopy of OxD}
For the deuterated species, OxD, the IR2015 spectrum\cite{wolke:2015}
was used to further validate the TL-PES and to provide spectroscopic
assignments. First, the measurements were compared with the VPT2
calculations, see Figure \ref{fig:oxod_vpt2}. Again, for the framework
modes (bands {\bf a} to {\bf f} below 2000 cm$^{-1}$) excellent
agreement is found, see Table ~\ref{sitab:oxd_assign} for numerical
values and assignments. The spectral patterns related to the deuterium
motion, which are located above 2000 cm$^{-1}$ are also very well
captured, in particular finer and low-intensity peaks around 3000
cm$^{-1}$ and above. However, there is a prominent pattern between
1860 and 1980 cm$^{-1}$ (blue shaded area) which is largely empty for
the VPT2 calculations. To further explore the spectroscopy in this
frequency range a perturbed minimum energy structure was generated
with a bond length standard deviation of 0.01 \AA\/. The result of the
VPT2 calculations were compared with the VPT2 calculation for the
minimum energy structure (red) and experimental data (brown), see
Figure \ref{sifig:oxod_vpt2_perturbed}. It is found that perturbation
away from the minimum energy geometry leads to a shift in the
OD-stretch frequency, consistent with the findings for OxH.\\

\begin{figure}[h!]
\centering
\includegraphics[width=1.0\textwidth]{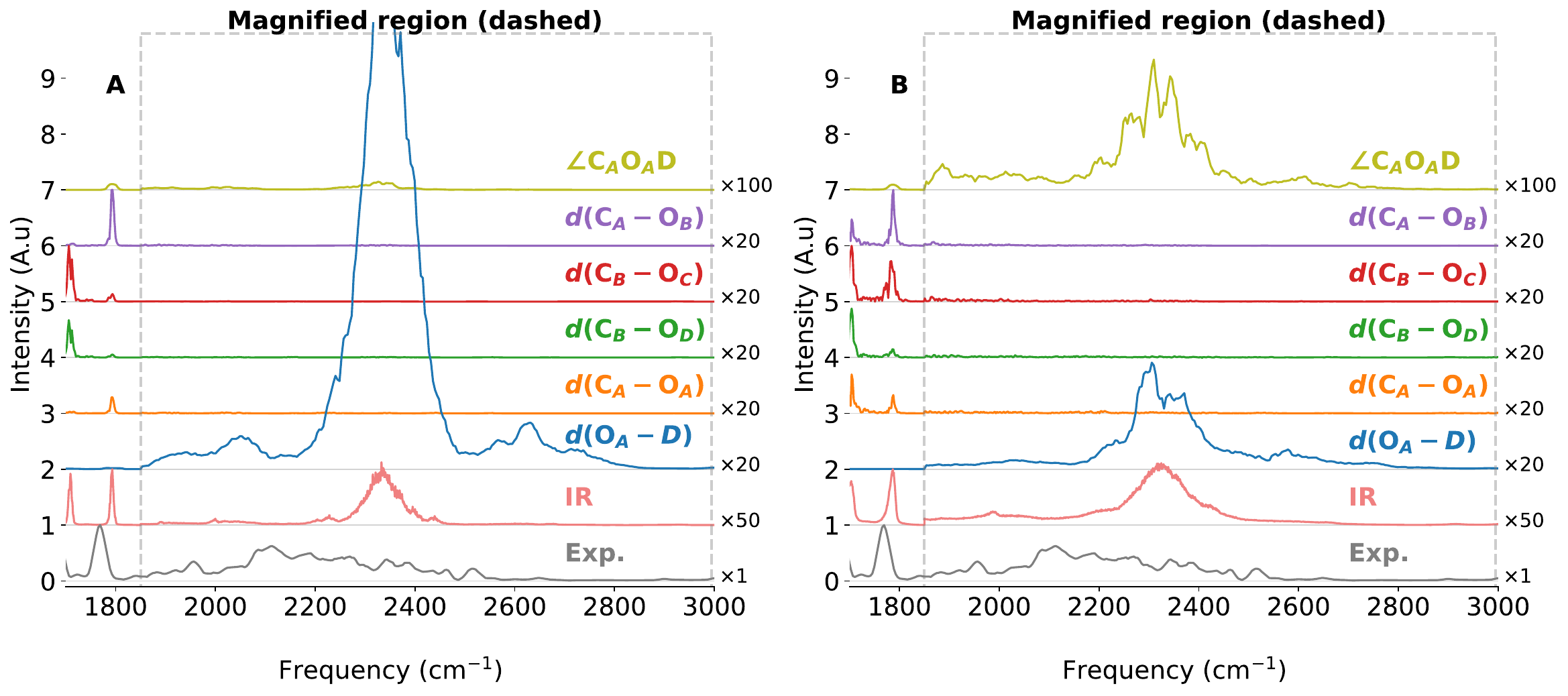}
\caption{Power Spectra and IR spectra for OxD (high-frequency
  region). Panel A: without deuterium transfer, Panel B: with
  deuterium transfer.}
\label{fig:oxd-spec}
\end{figure}

\noindent
Next, the infrared spectra were obtained from MD simulations at 600 K
together with power spectra for important internal degrees of
freedom. Again, trajectories without and with D-transfer were analyzed
separately, see Figures \ref{fig:oxd-spec}A and B. The OD-stretch
vibration likely corresponds to the peak observed at 2110 cm$^{-1}$ in
the experimental spectrum, whereas in the spectrum computed from MD
simulations, this feature appears at 2320 cm$^{-1}$. The shift of
approximately --200 cm$^{-1}$ for OxD is consistent with the
isotope-corrected shift observed for OxH ($-260/\sqrt{2} = -185$
cm$^{-1}$). The maximum peak position at 2110 cm$^{-1}$ in the
experimental spectrum of OxD, similar to that in OxH, is associated
with motions along the O$_{\rm A}$D-stretch and C$_{\rm A}$O$_{\rm
  A}$D-bending motions, see olive and blue traces Figure
\ref{fig:oxd-spec}B.\\

\subsection{H-Transfer and Tunneling Splittings}
First, the classical "over-the-barrier" H-transfer process was
considered from equilibrium MD simulations. Running classical dynamics
simulations at different temperatures, the number of barrier crossings
was determined from which the H-transfer rate can be obtained. For
this 500 MD simulations, 200 ps each (aggregate of 100 ns) were
carried out at temperatures $T \in [50, 900]$ using ASE. These
simulations yield proton transfer (PT) rates 0, 4.8, 131.4 and 339.0
ns$^{-1}$ at $T = 50, 300, 600$, and 900 K, respectively. Using an
Arrhenius expression $k(T) = Ae^{-\frac{E_a}{RT}}$ one obtains $E_a =
3.84$ kcal/mol, and $A = 3057.3$ ns$^{-1}$. This compares and is
consistent with a barrier height of 3.35 kcal/mol from the
CCSD(T)/aug-cc-pVTZ calculations.\\

\noindent
For a first estimation of the tunneling splittings, sRPI theory was
employed together with the single-precision PhysNet representation of
the PES. Such a treatment is considered appropriate for cases in which
anharmonic effects perpendicular to the instanton path are less
pronounced and the barrier for H-transfer is not particularly low. All
leading-order splitting calculations in the present work were carried
out with up to $N=4096$ beads at an effective temperature of $T_{\rm
  eff} = \hbar/k_{\rm B} \tau = 25$~K. Using the MP2-level PES,
featuring a barrier height $E_{\rm B} = 2.355$ kcal/mol, $\Delta_{\rm
  H} = 116.5$ cm$^{-1}$ was obtained. As for the higher-level TL-PES
$E_{\rm B}$ increases to 3.351 kcal/mol, the tunneling splitting
decreases to $\Delta_{\rm H} = 36.0$ cm$^{-1}$. The numerical
convergence of the leading-order tunneling splittings as a function of
$N$ is reported in Table~\ref{sitab:convergence}. \\

\noindent
One limitation of sRPI for determining tunnelling splittings is that
fluctuations around the instanton are treated within a harmonic
approximation (i.e. the Hessians are used). As a remedy, a
perturbatively corrected approach was developed which is based on
third and fourth order derivatives of the PES along the instanton
path.\cite{lawrence2023perturbatively} For numerical stability, pcRPI
requires highly accurate higher-order derivatives. Within the context
of machine learning-based PESs it is common to use single-precision
arithmetics for training and evaluation, which was found to be
insufficient for pcRPI.\cite{kaeser2024numerical} Hence, a
double-precision TL-PES was obtained from a retrained MP2 PhysNet
model also using double precision arithmetics throughout. The higher
order instanton calculations were carried out at $T_{\rm eff} = 25$~K
and with $N=1024$ beads as pcRPI is considerably more memory intensive
than sRPI, and the perturbative correction was found to converge
significantly more rapidly than the leading-order
term.\cite{MM.tl:2025} The correction factor for H-transfer was
$c_{\rm PC} = 0.98$ within pcRPI which leads to $\Delta^{\rm
  pcRPI}_{\rm H} = 35$ cm$^{-1}$. This is 1 cm$^{-1}$ smaller than the
tunneling splitting of 36.0 cm$^{-1}$ determined on the single
precision TL-PES. Assuming a conservative error estimate of 20 \% for
sRPI theory applied to the computed tunneling splitting using the
TL-PES brackets the prediction for H-transfer splitting to around
$\Delta_{\rm H} \in [28,42]$ cm$^{-1}$. More accurate predictions will
likely require further refinement of the TL-PES, with a specific focus
on tunneling splittings\cite{MM.tlma:2022,MM.tl:2025} which are very
sensitive to the shape of the PES, in particular around the global
minimum and the barrier to H-transfer.\\

\noindent
Earlier, recent work using transfer-learned PESs to the CCSD(T) level
of theory for malonaldehyde and tropolone showed excellent agreement
with measured tunneling
splittings.\cite{MM.tlma:2022,kaeser2024numerical,MM.tl:2025} For
malonaldehyde the measured\cite{baba:1999} splitting $\Delta_{\rm
  H}^{\rm expt.} = 21.6$ cm$^{-1}$ compares with a computed sRPI
(pcRPI) $\Delta_{\rm H}^{\rm TL} = 25.3$ (22.1) cm$^{-1}$ on a PES
transfer learned to the CCSD(T) level of theory and featuring a
barrier height of $E_{\rm B} = 3.89$ kcal/mol. For
tropolone,\cite{tanaka:1999} on the other hand, $\Delta_{\rm H}^{\rm
  expt.} = 0.97$ cm$^{-1}$ is considerably smaller because $E_{\rm B}
= 6.62$ kcal/mol at the CCSD(T) level of theory and pcRPI $\Delta_{\rm
  H}^{\rm theo.} = 0.94$ cm$^{-1}$. Hence increasing the barrier by
ca. 2 kcal/mol reduces $\Delta_{\rm H}$ by approximately 1 order of
magnitude. This is consistent with an increase to $\Delta_{\rm H} =
35.0$ cm$^{-1}$ for oxalate because the barrier to HT decreases by 0.5
kcal/mol (to $E_{\rm B} = 3.35$ kcal/mol) compared with
malonaldehyde.\\

\section{Summary and Conclusion}
The present work provides a molecularly resolved picture for the
dynamics, spectroscopy and H/D-transfer dynamics in the oxalate
anion. VPT2 calculations and finite-temperature MD simulations using a
CCSD(T)-quality PES provide very realistic spectral patterns for the
framework modes (below 2000 cm$^{-1}$), including the prediction of
the $\nu_5 + \nu_{10}$ combination band (feature {\bf e$_1$}) at 1652
cm$^{-1}$ from VPT2 calculations which is clearly defined (observed at
1666 cm$^{-1}$) in the new measurements (IR2025, see Figure
\ref{sifig:zoomed_figure3}).\\

\noindent
A particular focus was on exploring the spectral signatures between
2400 and 3200 cm$^{-1}$ for OxH and 1900 to 2500 cm$^{-1}$ for OxD,
respectively. The VPT2 calculations find a high density of states in
these frequency ranges and slightly perturbing the structure away from
the minimum energy configuration further increases the density of
vibrational signatures. These calculations together with the
finite-temperature MD simulations identify a center-frequency at $\sim
2940$ and $\sim 2110$ cm$^{-1}$ as the main H-/D-transfer
mode. However, unequivocal assignment of one feature to a single
motional pattern is not meaningful because the fundamentals,
combination and hot bands are heavily mixed. \\

\noindent
Consistent with earlier work,\cite{wolke:2015} it is found here that
the COH angle strongly couples to the OH stretching frequencies. This
is particularly apparent for trajectories that feature H-transfer. The
VPT2 calculations combined with MD simulations at the highest level of
theory reported in the literature provided assignments of the
vibrational modes and correct previous mode descriptions. The mismatch
in assignment with the one made for IR2015 is mostly related to using
a comparatively low level of quantum chemical theory
(B3LYP/6-311++G(d,p)). Using a reactive force field, the effective
barrier height for H-transfer was also inferred from matching the
measured IR2015 spectrum with that computed from the
dipole-autocorrelation function from finite-temperature MD
simulations. An effective barrier height of 4.2 kcal/mol was found to
best reproduce the IR2015 in the region of the H-transfer mode. This
compares with an electronic barrier height of 3.35 kcal/mol from the
present work and 3.4 kcal/mol from previous CCSD(T)/aug-cc-pVTZ
calculations.\cite{boutwell:2022} However, the electronic and
effective barrier heights describe different physical quantities and
thus can not be compared directly. The electronic $E_{\rm B}$
represents the pure potential energy difference between reactant and
the transition state whereas the effective barrier height scales the
PES along a single (reaction) coordinate (here the HT-motion), depends
on temperature, and accounts for zero-point effects.\\

\noindent
Tunneling splittings for OxH have been determined with RPI theory.  A
rather conservative estimate for the tunneling splitting is
$\Delta_{\rm H}^{\rm sRPI} \in [28,42]$ cm$^{-1}$ which, however, is
likely to require improvements of the TL-PES for quantitative
predictions of potential future experiments.  Nevertheless, the
predicted range offers a reference for future spectroscopic efforts
and experimental confirmation would not only validate the RPI
framework but also aid in the iterative improvement of the PES guided
by experiments as has been recently done for small ionic complexes
through morphing.\cite{MM.morph:2024,MM.morph:1999,gazdy:1991}\\

\noindent
In summary, an updated measured infrared spectrum for the oxalate
anion was discussed in the context of state-of-the art VPT2
calculations and MD simulations using a transfer-learned CCSD(T)-level
PES. Assignments were made based on inspecting the VPT2 vibrations and
the power spectra. In the region of the framework modes considerable
coupling between the different degrees of freedom was found. The
agreement between theory and experiment and the fact that a specific
framework mode was correctly located by the VPT2 calculations
highlights the predictive power of the theoretical approaches, whereby
vibrational motion can be reliably related to observed spectral
features. Examples such as OxH and OxD are particularly valuable as
they allow computer-based methods to approach "prediction mode",
finding the "right answer for the right reason".\\

\section*{Supporting Information}
Figures \ref{sifig:errcorr} and \ref{sifig:errcorr_ccsd} present the
performance of PhysNet models trained at different levels of
theory. Figures \ref{sifig:ir_hoxa_physnet} to
\ref{sifig:oxod_vpt2_perturbed} provide supplementary analyses and
spectral data that support and extend the results discussed in the
main manuscript. Tables \ref{sitab:harmfreq} to \ref{sitab:oxd_assign}
report a detailed comparison between the vibrational frequencies
predicted by the models and those obtained from reference quantum
chemical methods and experimental measurements, including peak
assignments. Table \ref{sitab:convergence} summarizes the calculated
tunneling splittings together with their convergence.

\section*{Data Availability}
The PhysNet codes and input files for the MP2 and
CCSD(T) PESs are available from
\url{https://github.com/MMunibas/oxalate/}.

\section*{Acknowledgment}
The authors gratefully acknowledge financial support from the Swiss
National Science Foundation through grants $200020\_219779$ (MM),
$200021\_215088$ (MM), the University of Basel (MM). MAJ gratefully
acknowledges the Department of Energy through the condensed phase and
interfacial molecular science (CPIMS) program under grants
DE-SC0021012. ELB was supported by the National Institutes of Health
(NIH) Biophysical Training grant no. 5T32GM149438-02.\\

\section*{Conflict of Interest}
The authors declare no conflict of interest.

\bibliography{refs}

\clearpage

\renewcommand{\thetable}{S\arabic{table}}
\renewcommand{\thefigure}{S\arabic{figure}}
\renewcommand{\theequation}{S\arabic{equation}}
\renewcommand{\thesection}{S\arabic{section}}
\setcounter{figure}{0}  
\setcounter{section}{0}  
\setcounter{table}{0}

\newpage

\noindent
{\bf SUPPORTING INFORMATION: Dynamics of Protonated Oxalate from
  Machine-Learned Simulations and Experiment: Infrared Signatures,
  Proton Transfer Dynamics and Tunneling Splittings}

\begin{figure}[h!]
\centering
\includegraphics[width=0.7\textwidth]{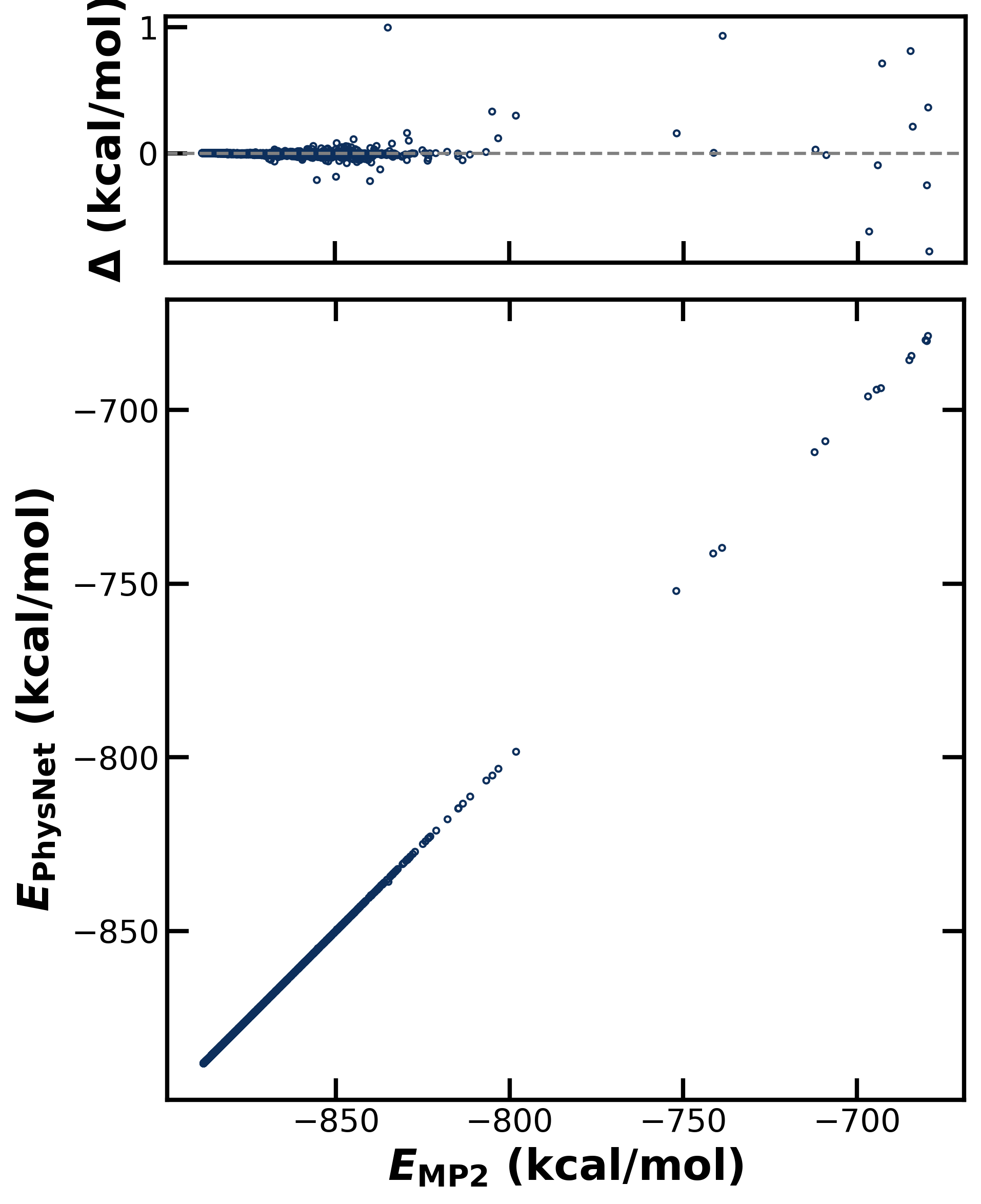}
\caption{The main panel shows the out of sample performance of the PhysNet
  model. The PhysNet model is trained on energies, forces and dipole
  moments determined at the MP2/aug-cc-pVTZ level of theory.
  Roughly $10~\%$ (2220) served as test set and were not used during training. 
  The top panel reports $\Delta = E_{\rm Ref.} - E_{\rm PhysNet}$. Note that the
  high error structures correspond to unphysical structures sampled during the 
  active learning cycle (e.g. the hydrogen atom resides between the two carbons).}
\label{sifig:errcorr}
\end{figure}

\begin{figure}[h!]
\centering
\includegraphics[width=0.7\textwidth]{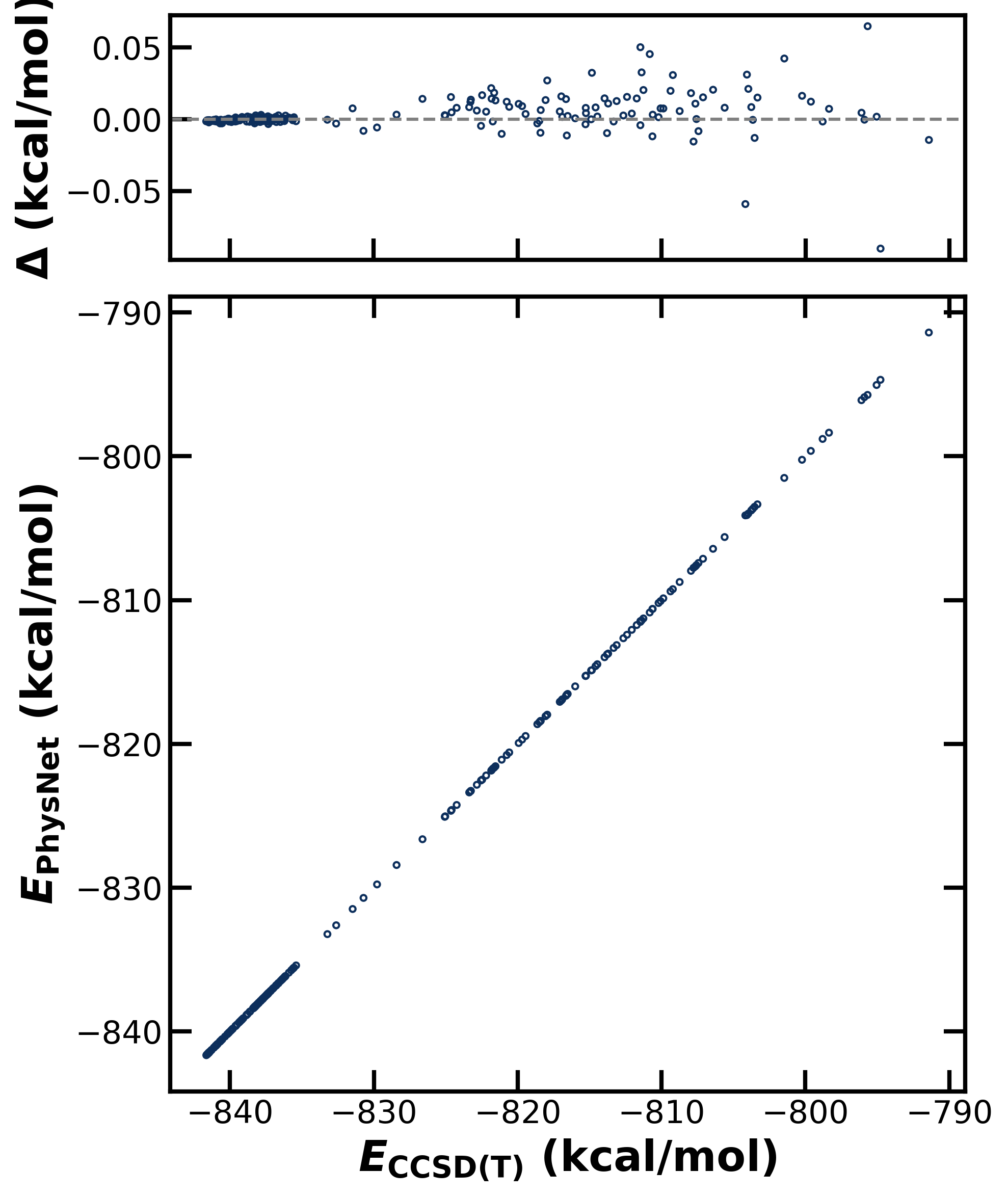}
\caption{The main panel shows the out of sample performance of the
  transfer-learned PhysNet model. The PhysNet model is
  transfer-learned on energies, forces and dipole moments determined
  at the CCSD(T)/aug-cc-pvtz level of theory for a total 2688
  structures.  270 structures ($10~\%$) served as test set and were
  not used during training.  The top panel reports $\Delta = E_{\rm
    Ref.} - E_{\rm PhysNet}$.}
\label{sifig:errcorr_ccsd}
\end{figure}

\begin{figure}[h!]
\centering \includegraphics[width=1.0\textwidth]{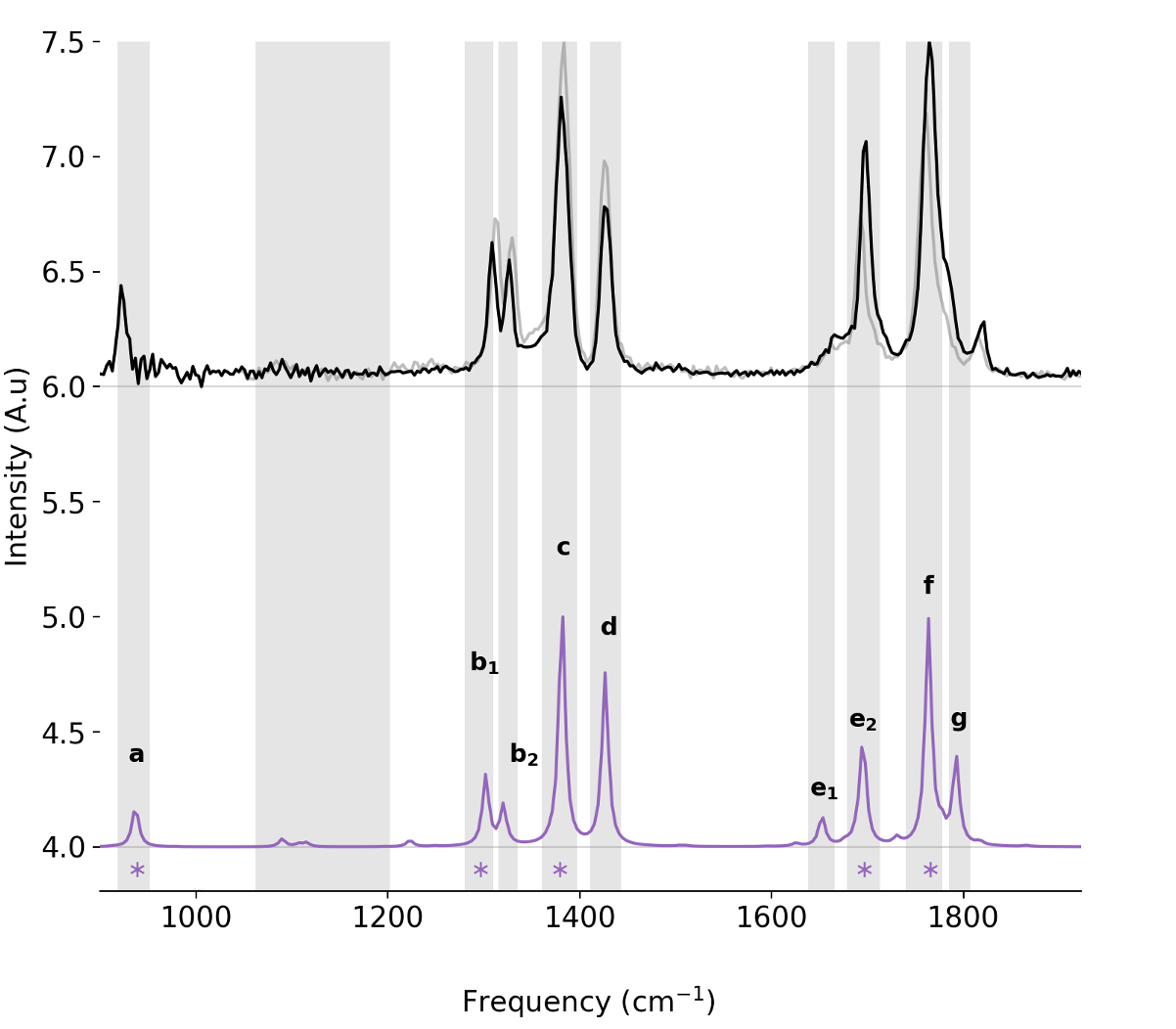}
\caption{Zoom-in of the two top traces in Figure
  \ref{fig:ir-sim}. Upper traces show IR2025 (black) and
  IR2015\cite{wolke:2015} (grey) and bottom trace reports the VPT2
  calculation using the ML-PES at the CCSD(T) level of theory.}
\label{sifig:zoomed_figure3}
\end{figure}

\begin{figure}[h!]
\centering
\includegraphics[width=1.0\textwidth]{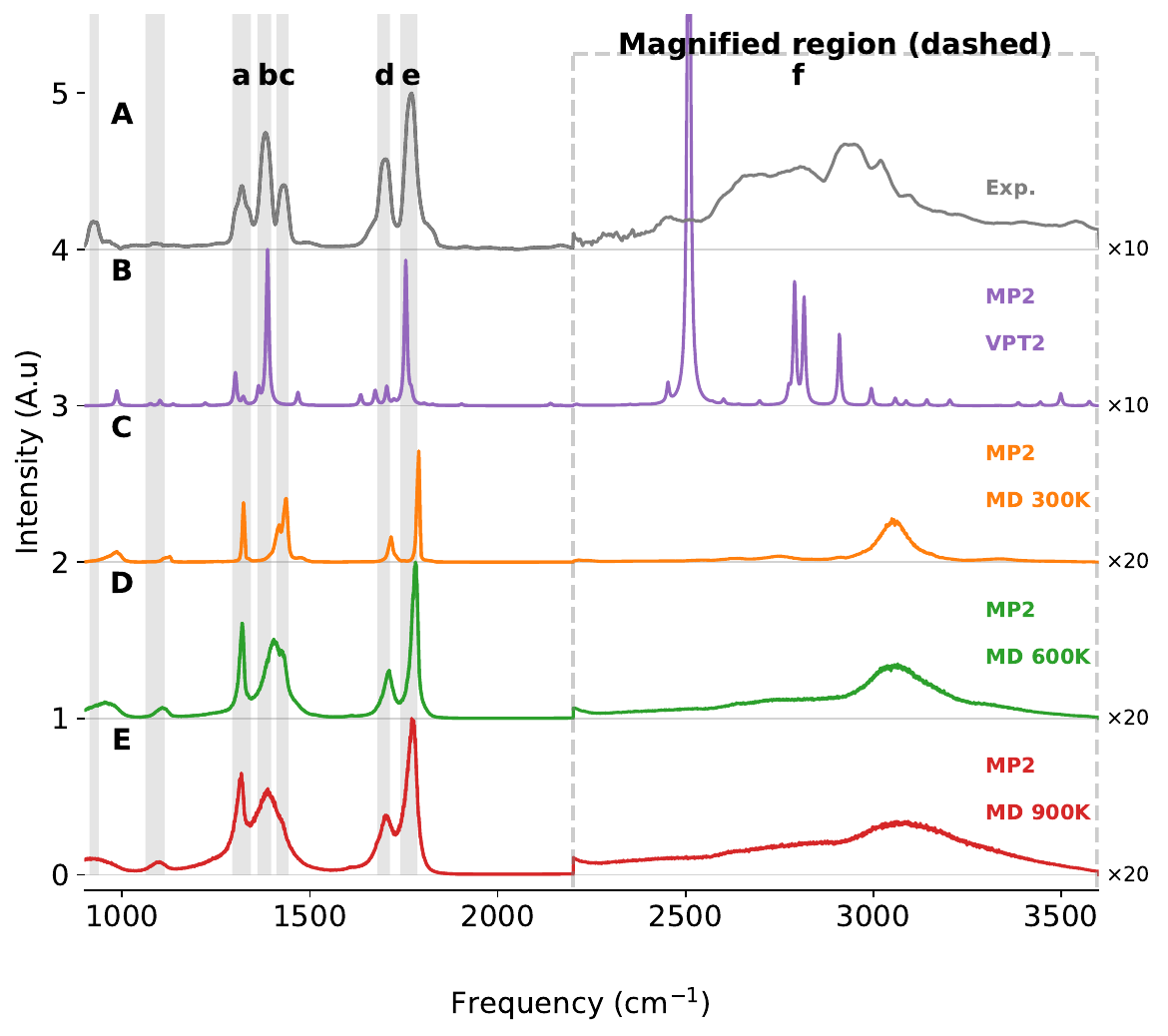}
\caption{Infrared spectra of OxH using the MP2 ML-PES. The upper trace
  is the experimentally determined (H$_2$-tagged) gas phase spectrum
  IR2025. The following four computed spectra (Panels B-E) are
  obtained from second-order vibrational perturbation theory
  (VPT2)\cite{barone:2010} and from the Fourier transform of the
  dipole-dipole autocorrelation function.}
\label{sifig:ir_hoxa_physnet}
\end{figure}

\begin{figure}[h!]
\centering \includegraphics[width=1.0\textwidth]{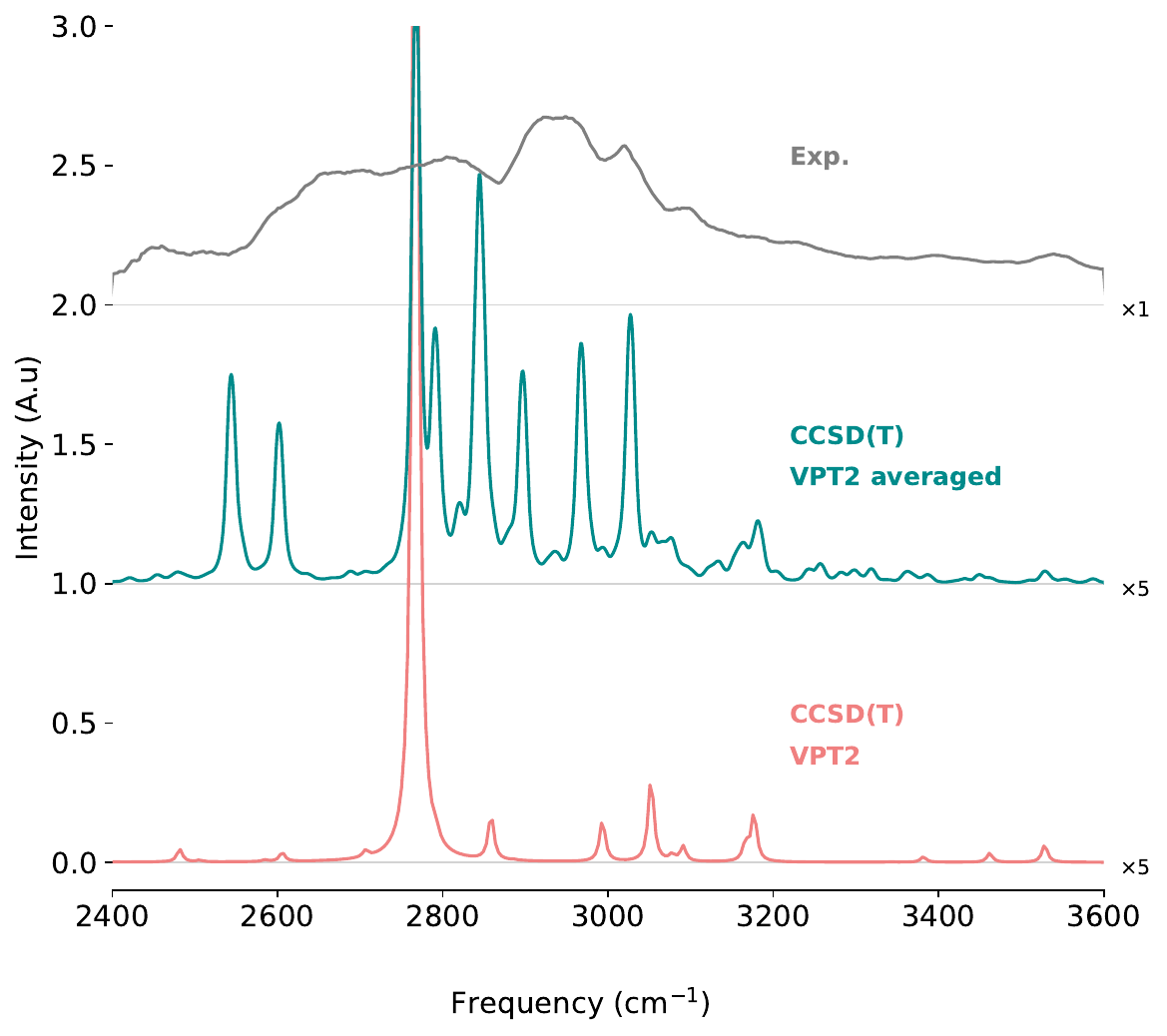}
\caption{Boltzmann-averaged VPT2 spectrum for OxH obtained from 9
  rattled geometries and minimum energy geometry using the
  transfer-learned PhysNet model of CCSD(T)/aug-cc-pvtz quality. The
  top trace reports the experimentally determined (H$_2$-tagged) gas
  phase spectrum IR2025 and two bottom trace is the computed spectrum
  from second-order vibrational perturbation theory (VPT2) for
  averaged ensemble (dark cyan) and minimum energy (light coral).}
\label{sifig:oxoh_vpt2_averaged}
\end{figure}

\begin{figure}[h!]
\centering
\includegraphics[width=1.0\textwidth]{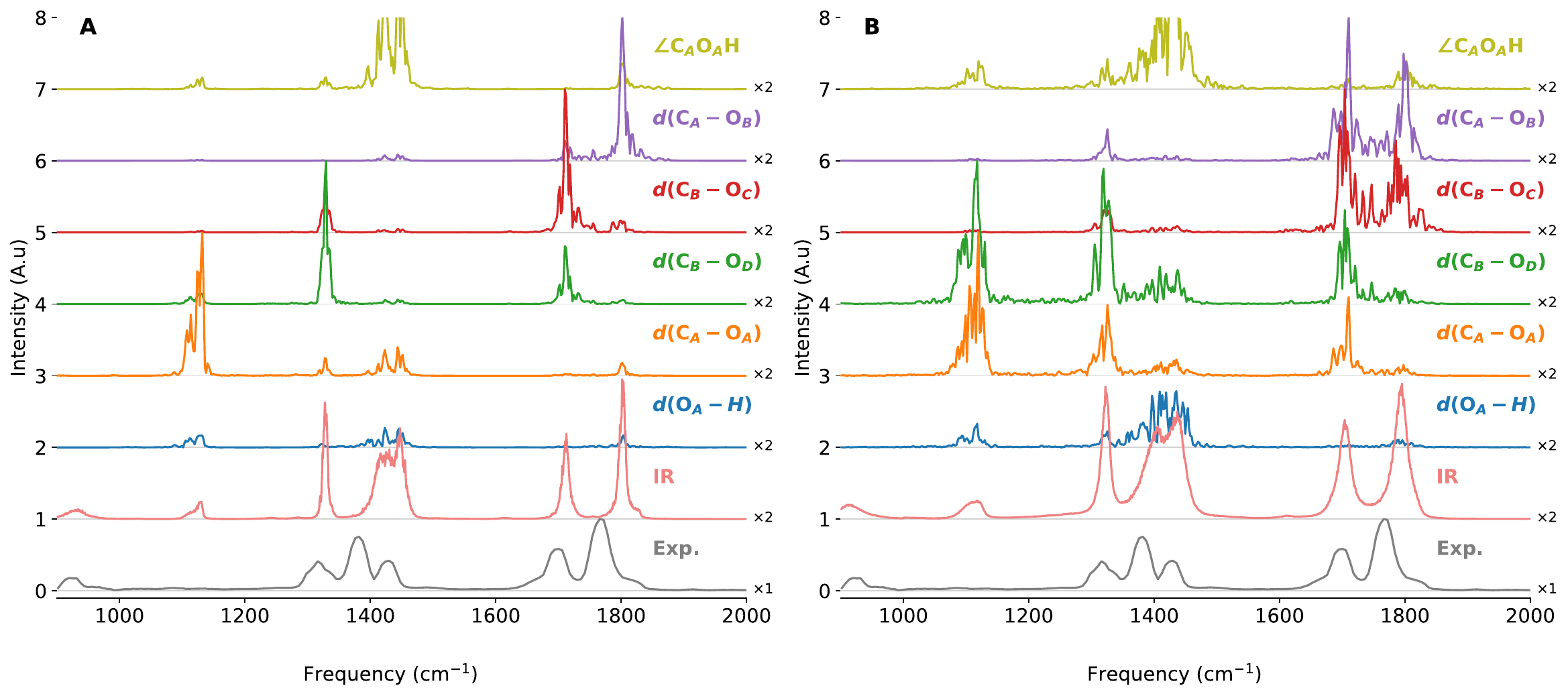}
\caption{The low-frequency region of the OxH power and IR spectra from
  MD simulations at 600K using the PhysNet (CCSD(T)) model. Panel A:
  without proton transfer, Panel B: with proton transfer.}
\label{sifig:ps_600K_OxOH_low}
\end{figure}

\begin{figure}[h!]
\centering
\includegraphics[width=1.0\textwidth]{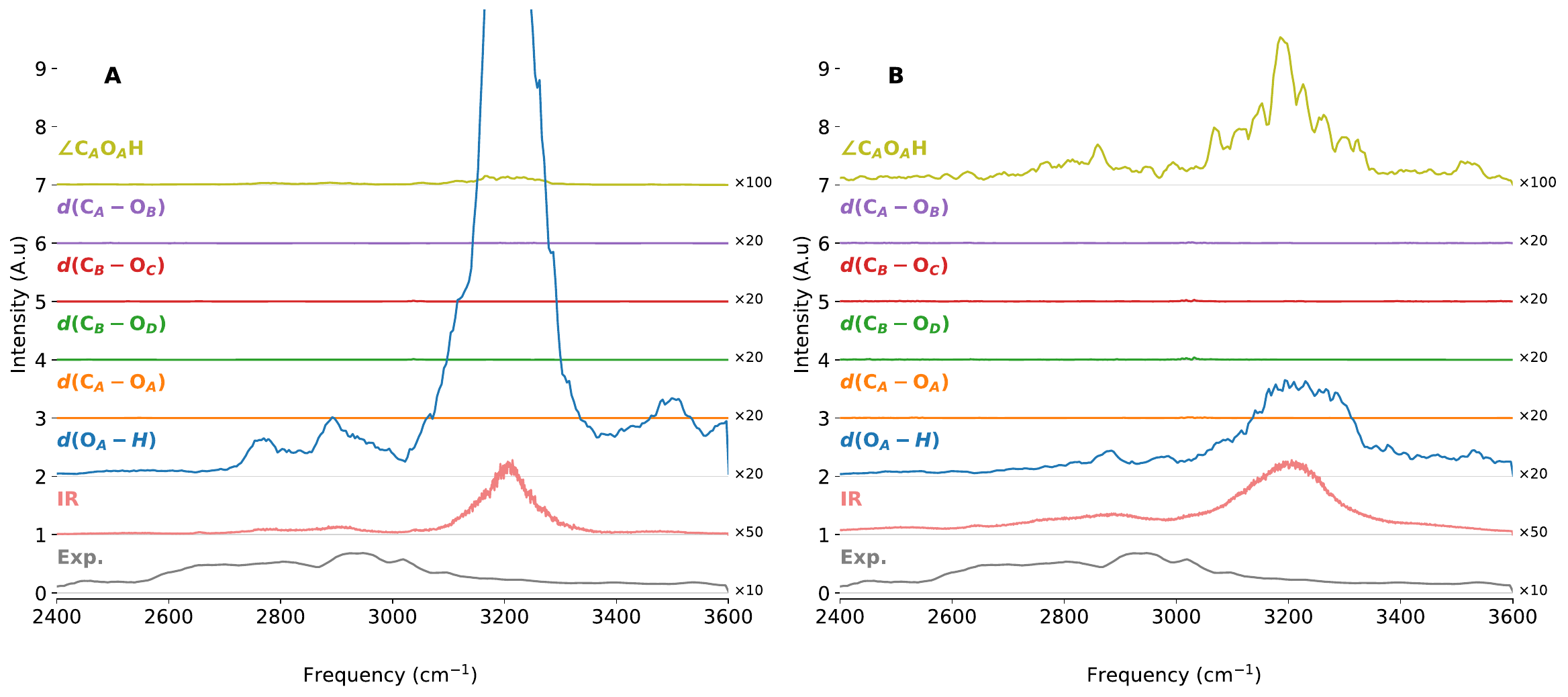}
\caption{The high-frequency region of the OxH power and IR spectra
  from MD simulations at 600K using the PhysNet (CCSD(T)) model. Panel
  A: without proton transfer, Panel B: with proton transfer.}
\label{sifig:ps_600K_OxOH_high}
\end{figure}

\begin{figure}[h!]
\centering
\includegraphics[width=1.0\textwidth]{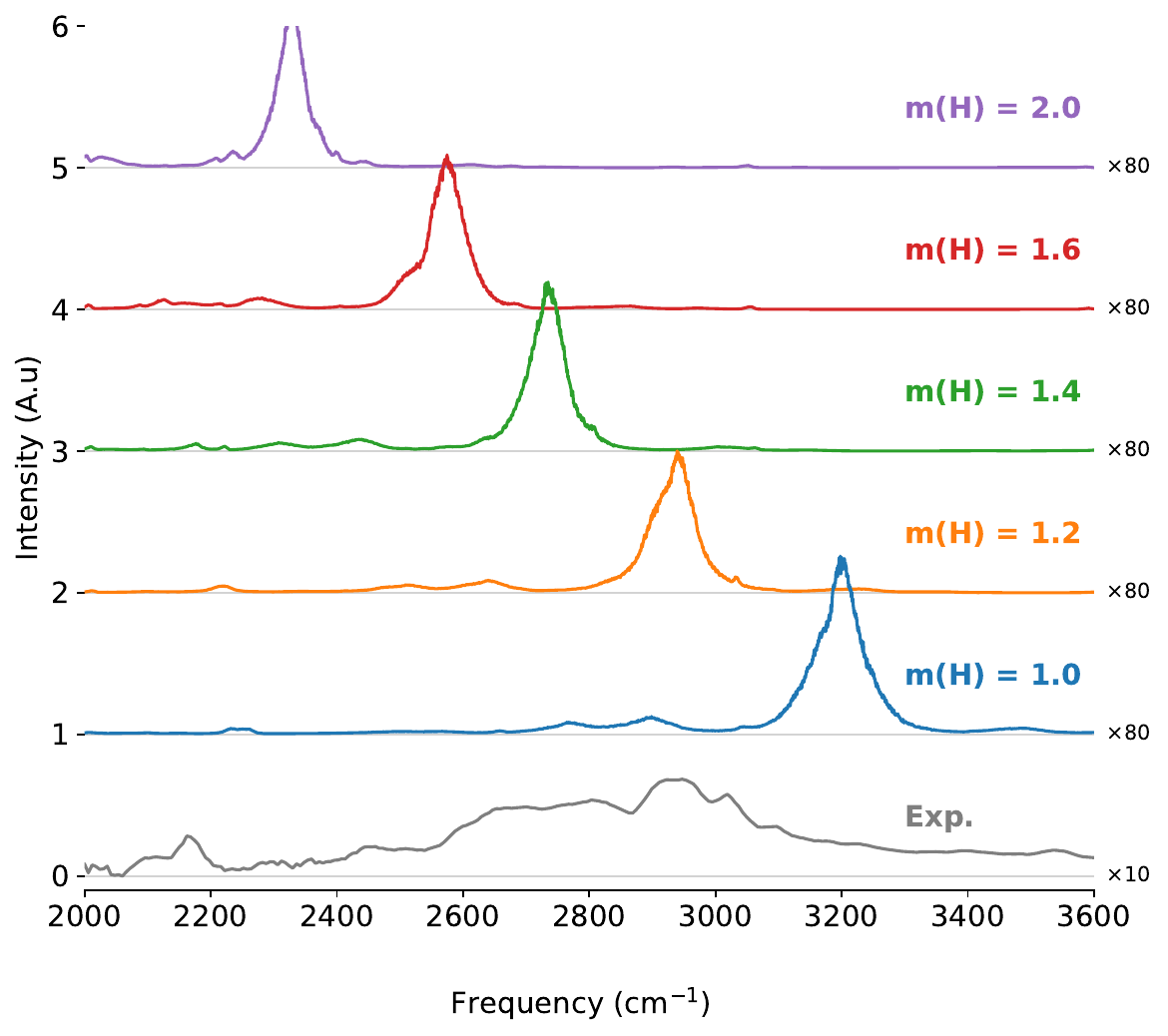}
\caption{IR spectra at 300K from MD simulations with masses of the
  H-atom ranging from $m_{\rm H}$ to $m_{\rm D}$ using the PhysNet
  (CCSD(T)) ML-PES.}
\label{sifig:masses}
\end{figure}

\begin{figure}[h!]
\centering \includegraphics[width=1.0\textwidth]{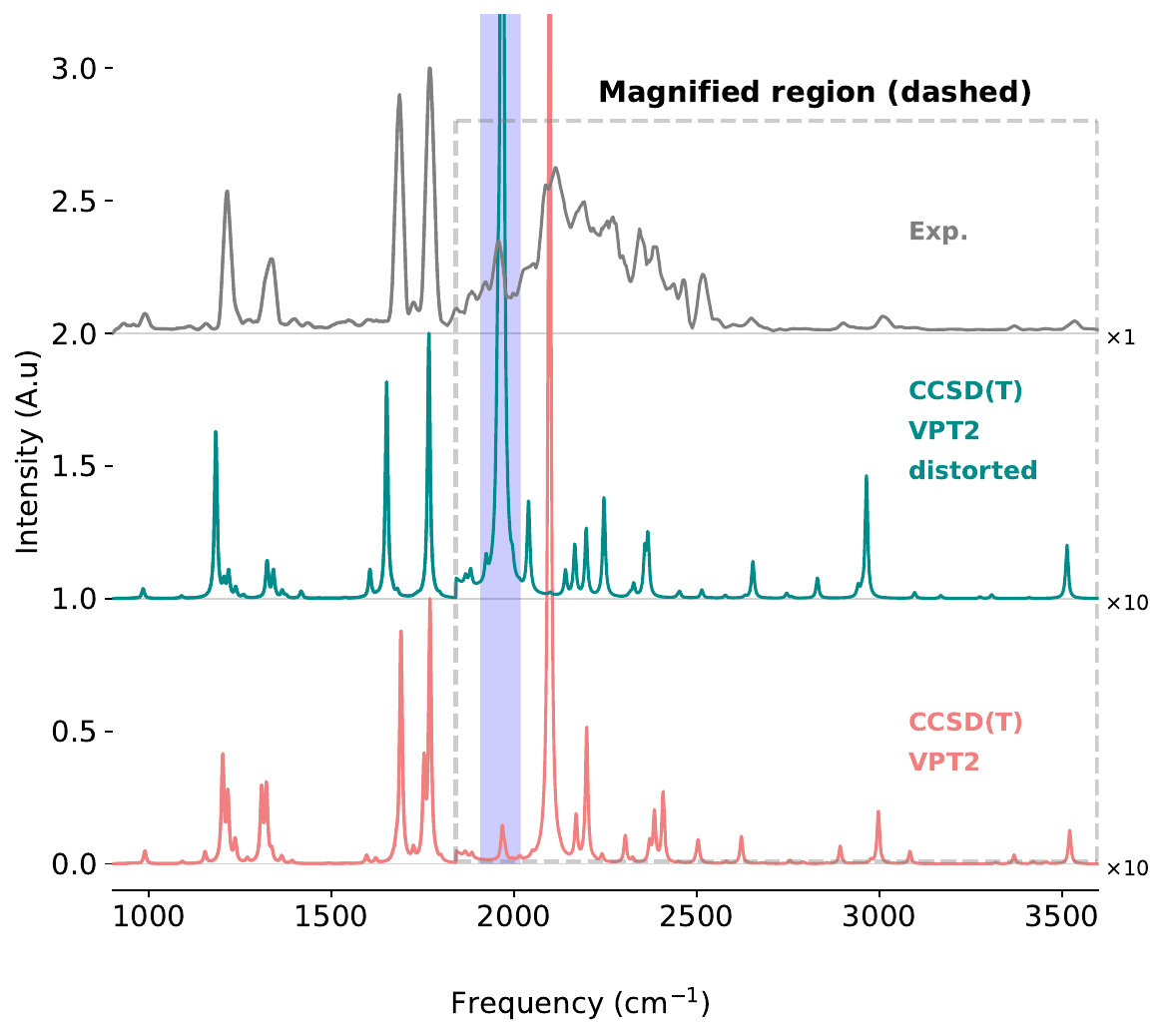}
\caption{The IR spectra for OxD. The top trace reports the
  experimentally determined (H$_2$-tagged) gas phase spectrum IR2015
  and the two bottom traces are the computed VPT2- spectra. The
  computations use the PhysNet model trained at CCSD(T)/aug-cc-pVTZ
  level of theory and the minimum energy structure (light coral) and a
  slightly distorted structure (dark cyan).}
\label{sifig:oxod_vpt2_perturbed}
\end{figure}

\clearpage

\begin{table}[]
\begin{tabular}{rrrrrrr}
 & \multicolumn{3}{c}{\textbf{Min}} & \multicolumn{3}{c}{\textbf{TS}}\\\toprule
 & \textbf{PhysNet} & \textbf{MP2} & $|\Delta|$&  \textbf{PhysNet} & \textbf{MP2}&$|\Delta|$ \\\midrule
1      & 109.63  & 109.65  & 0.02 & 993.74i & 993.67i & 0.07 \\
2      & 299.29  & 299.25  & 0.04 & 143.18  & 143.16  & 0.02 \\
3      & 434.22  & 434.31  & 0.09 & 334.65  & 334.61  & 0.04 \\
4      & 488.16  & 488.13  & 0.03 & 492.19  & 492.43  & 0.24 \\
5      & 572.13  & 572.13  & 0.00 & 605.50  & 605.55  & 0.05 \\
6      & 699.97  & 699.96  & 0.01 & 711.16  & 711.15  & 0.01 \\
7      & 830.24  & 830.30  & 0.06 & 753.88  & 753.80  & 0.08 \\
8      & 850.41  & 850.52  & 0.11 & 853.21  & 853.33  & 0.12 \\
9      & 1004.39 & 1004.59 & 0.20 & 863.82  & 864.02  & 0.20 \\
10     & 1134.95 & 1134.99 & 0.04 & 1290.59 & 1290.56 & 0.03 \\
11     & 1328.41 & 1328.54 & 0.13 & 1301.42 & 1301.50 & 0.08 \\
12     & 1448.93 & 1449.07 & 0.14 & 1320.64 & 1320.88 & 0.24 \\
13     & 1726.28 & 1726.47 & 0.19 & 1738.65 & 1738.70 & 0.05 \\
14     & 1798.97 & 1799.09 & 0.12 & 1782.98 & 1783.17 & 0.19 \\
15     & 3067.27 & 3067.09 & 0.18 & 2081.24 & 2080.99 & 0.25 \\\midrule
\textbf{MAE}& && \textbf{0.09}  &&& \textbf{0.11}  \\\bottomrule
\end{tabular}
\caption{Harmonic frequencies (all values in cm$^{-1}$), obtained from
  the Hessian matrix ($H=\partial^{2} E / \partial
  \boldsymbol{r}^{2}$) using the PhysNet PES trained on
  MP2/aug-cc-pVTZ reference data compared with \textit{ab initio}
  calculations at the MP2/aug-cc-pVTZ level of theory. Results for the
  minimum energy (Min) and transition state (TS) structures are
  reported. The last row reports the MAE between the two different
  calculations.}
\label{sitab:harmfreq}
\end{table}

\begin{table}[]
\begin{tabular}{rrrrrrr}
 & \multicolumn{3}{c}{\textbf{Min}} & \multicolumn{3}{c}{\textbf{TS}}\\\toprule
 & \textbf{PhysNet} & \textbf{CCSD(T)} & $|\Delta|$&  \textbf{PhysNet} & \textbf{CCSD(T)}&$|\Delta|$ \\\midrule
1	&	104.35	&	104.16	&	0.19	&	1141.98i	&	1143.01i	&	1.03	\\
2	&	300.33	&	300.53	&	0.20	&	141.9	&	142.08	&	0.18	\\
3	&	434.36	&	434.85	&	0.49	&	333.57	&	333.55	&	0.02	\\
4	&	489.59	&	488.55	&	1.04	&	494.51	&	492.74	&	1.77	\\
5	&	565.59	&	565.63	&	0.04	&	602.15	&	602.51	&	0.36	\\
6	&	700.52	&	700.53	&	0.01	&	715.54	&	715.09	&	0.45	\\
7	&	829.31	&	829.31	&	0.00	&	750.43	&	750.74	&	0.31	\\
8	&	848.72	&	849.26	&	0.54	&	852.04	&	852.74	&	0.70	\\
9	&	967.49	&	968.64	&	1.15	&	866.12	&	865.8	&	0.32	\\
10	&	1140.14	&	1140.19	&	0.05	&	1296.59	&	1296.66	&	0.07	\\
11	&	1334.78	&	1335.09	&	0.31	&	1299.53	&	1299.88	&	0.35	\\
12	&	1455.55	&	1455.66	&	0.11	&	1321.83	&	1322.91	&	1.08	\\
13	&	1726.48	&	1726.89	&	0.41	&	1748.49	&	1748.71	&	0.22	\\
14	&	1816.62	&	1816.52	&	0.10	&	1789.66	&	1790.83	&	1.17	\\
15	&	3204.38	&	3203.91	&	0.47	&	2100.79	&	2101.26	&	0.47	\\\midrule
\textbf{MAE}& && \textbf{0.34} & & & \textbf{0.57} \\\bottomrule
\end{tabular}
\caption{Harmonic frequencies (all values in cm$^{-1}$), obtained from
  the Hessian matrix ($H=\partial^{2} E / \partial
  \boldsymbol{r}^{2}$) using the PhysNet PES transfer-learned using
  CCSD(T) /aug-cc-pVTZ reference data compared with \textit{ab initio}
  calculations at the CCSD(T)/aug-cc-pVTZ level of theory. Results for
  the minimum energy (Min) and transition state (TS) structures are
  reported. The last row reports the MAE between the two different
  calculations.}
\label{sitab:harmfreq_tl}
\end{table}

\begin{table}[htbp]
\caption{Harmonic and anharmonic Frequencies (VPT2) using the PhysNet
  (CCSD(T)) ML-PES and experimentally measured ($\pm 1$ cm$^{-1}$)
  vibrational transitions of OxH(H$_2$)$_2$. $^\dagger$ weak ($\pm 5$
  cm$^{-1}$). The RMSD (cm$^{-1}$) between VPT2 and experiment for
  fundamentals is 5.5 cm$^{-1}$, and for all modes below 1822
  cm$^{-1}$ it is 18.2 cm$^{-1}$. Mode $\nu_9$ was
  originally\cite{wolke:2015} labelled as $\nu_{12}$ and assigned to
  both C out-of-plane and in-phase motion. According to the VPT2
  calculations this is the OH out of plane motion. Mode descriptions:
  $\nu_1 \equiv$ Coupled asynchronous twisting of --C$_{\rm A}$O$_{\rm
    A}$O$_{\rm B}$ and --C$_{\rm B}$O$_{\rm C}$O$_{\rm D}$ groups,
  $\nu_2 \equiv$ Coupled asynchronous rocking of --C$_{\rm A}$O$_{\rm
    A}$O$_{\rm B}$ and --C$_{\rm B}$O$_{\rm C}$O$_{\rm D}$ groups,
  $\nu_3 \equiv$ --C$_{\rm A}$O$_{\rm A}$O$_{\rm B}$ bending, $\nu_4
  \equiv$ both C out-of-plane and in-phase, $\nu_{5} \equiv$ O$_{\rm
    A}$C$_{\rm A}$C$_{\rm B}$ scissoring, $\nu_{6} \equiv$ O$_{\rm
    C}$C$_{\rm B}$O$_{\rm D}$ scissoring, $\nu_{7} \equiv$ O$_{\rm
    C}$C$_{\rm B}$O$_{\rm D}$ scissoring (large amplitude), O$_{\rm
    A}$C$_{\rm A}$O$_{\rm B}$ scissoring and C$_{\rm A}$-C$_{\rm B}$
  stretching, $\nu_8 \equiv$ both C out-of-plane and out-of-phase,
  $\nu_9 \equiv$ O$_{\rm A}$H out-of-plane bending, $\nu_{10} \equiv$
  acid C$_{\rm A}$--O$_{\rm A}$ stretching, $\nu_{11} \equiv$
  --CO$_2^-$ symmetric stretching, $\nu_{12} \equiv$ O$_{\rm A}$H
  in-plane bending, $\nu_{13} \equiv$ --CO$_2^-$ asymmetric
  stretching, $\nu_{14} \equiv$ C$_{\rm A}$=O$_{\rm B}$ stretching,
  $\nu_{15} \equiv$ O$_{\rm A}$H stretching. See also Figure
  \ref{fig:ir-sim}.}
\begin{tabular}{ccccc}
\toprule
\textbf{Mode} & \textbf{Peak} & \textbf{Harmonic} & \textbf{Anharmonic} & \textbf{Experimental} \\
 &\textbf{label} & \textbf{Frequency (cm$^{-1}$)} &  \textbf{Frequency (cm$^{-1}$)} & \textbf{Transition (cm$^{-1}$)} \\
\midrule
$\nu_{9}$ & a & 967.58  & 936.56  & 928 \\
$\nu_{10}$ & & 1140.48  & 1089.69  & 1091$^\dagger$ \\
$\nu_3 +\nu_6$ & & ---  & 1107.17  & --- \\
$2\nu_5$ & & --- &  1114.80  & --- \\
$\nu_3+\nu_7$ & & ---  & 1222.92  & 1171(?) \\
$\nu_{11}$ & $\mathrm{b_1}$ & 1335.18 & 1301.92 & 1312\\
$\nu_4+\nu_8$ &$\mathrm{b_2}$ & --- & 1320.17 & 1327 \\
$\nu_{12}$ &  c  & 1455.77 & 1381.40  & 1381 \\
$\nu_4+\nu_9$& d & ---  & 1426.41 & 1429 \\
$\nu_5+\nu_{10}$& $\mathrm{e_1}$ & --- & 1652.48 & 1666 \\
$\nu_{13}$&$\mathrm{e_2}$ & 1727.10 & 1695.42 & 1696 \\
$\nu_{14}$& f & 1817.23  & 1763.54 & 1765 \\
$\nu_{6} + \nu_{10}$& g & ---  & 1792.3 & 1822 \\
$2\nu_{10}$ & & ---  & 2177.36  & --- \\
$\nu_{10}+\nu_{12}$ & & ---  & 2481.53 & 2457 \\
$2\nu_{11}$ & & ---  & 2605.59 & 2511 \\
$\nu_{15}$ & &  3204.67 & 2767.01 & 2600--3400 \\
$\nu_{10}+\nu_{14}$ & & --- & 2858.38 & --- \\
$\nu_{11}+\nu_{13}$ & & ---  & 2993.22 & --- \\
$\nu_2+\nu_{15}$ & & --- & 3052.01 & --- \\
$\nu_3+\nu_{15}$ & &  ---  & 3176.62 & --- \\
$2\nu_{14}$ & &  ---  & 3528.49 & 3528 \\
\bottomrule
\end{tabular}
\label{sitab:oxh_assign}
\end{table}

\begin{table}[htbp]
  \caption{Harmonic and anharmonic Frequencies (VPT2) using the
    PhysNet (CCSD(T)) ML-PES and experimentally\cite{wolke:2015}
    measured ($\pm 1$ cm$^{-1}$) vibrational transitions of
    OxD(H$_2$)$_2$. The RMSD (cm$^{-1}$) between VPT2 and experiment
    for fundamentals is 8.4 cm$^{-1}$, and for all modes below 1780
    cm$^{-1}$ it is 9.6 cm$^{-1}$. $^\dagger$ weak ($\pm 5$
    cm$^{-1}$). Mode descriptions: $\nu_3 \equiv$ --C$_{\rm A}$O$_{\rm
      A}$O$_{\rm B}$ scissoring, $\nu_4 \equiv$ both C out-of-plane
    and in-phase, $\nu_{5} \equiv$ O$_{\rm A}$C$_{\rm A}$C$_{\rm B}$
    scissoring, $\nu_{6} \equiv$ O$_{\rm A}$C$_{\rm A}$O$_{\rm B}$
    scissoring (large amplitude) and O$_{\rm C}$C$_{\rm B}$O$_{\rm D}$
    scissoring (small amplitude), $\nu_{7} \equiv$ O$_{\rm A}$D
    out-of-plane bending, $\nu_8 \equiv$ O$_{\rm C}$C$_{\rm B}$O$_{\rm
      D}$ scissoring (large amplitude) and O$_{\rm A}$C$_{\rm
      A}$O$_{\rm B}$ scissoring (small amplitude), $\nu_9 \equiv$ both
    C out-of-plane and out-of-phase, $\nu_{10} \equiv$ acid C$_{\rm
      A}$--O$_{\rm A}$ stretching, $\nu_{11} \equiv$ O$_{\rm A}$D
    in-plane bending, $\nu_{12} \equiv$ --CO$_2^-$ symmetric
    stretching, $\nu_{13} \equiv$ --CO$_2^-$ asymmetric stretching,
    $\nu_{14} \equiv$ C$_{\rm A}$=O$_{\rm B}$ stretching, $\nu_{15}
    \equiv$ O$_{\rm A}$D stretching. See also Figure
    \ref{fig:oxod_vpt2}.}
\begin{tabular}{ccccc}
\toprule
\textbf{Mode} & \textbf{Peak} & \textbf{Harmonic} & \textbf{Anharmonic} & \textbf{Experimental} \\
 &\textbf{label} & \textbf{Frequency (cm$^{-1}$)} &  \textbf{Frequency (cm$^{-1}$)} & \textbf{Transition (cm$^{-1}$)} \\
\midrule
$\nu_{10}$& a & 1024.16  & 989.15  & 985 \\
$\nu_4+\nu_7$& & ---  & 1153.78  & 1155$^\dagger$ \\
$\nu_{11}$& $\mathrm{b_1}$ & 1235.12  & 1202.17  & 1217 \\
$\nu_3 + \nu_8$& $\mathrm{b_2}$& ---  & 1216.32  & 1217 \\
$\nu_5+\nu_6$& c & ---  & 1236.50  & 1235$^\dagger$ \\
$\nu_4 + \nu_9$&$\mathrm{d_1}$ & --- & 1308.60 & ---\\
$\nu_{12}$&$\mathrm{d_2}$ & 1346.82 & 1321.76 & 1322 \\
$2\nu_7$& & --- & 1336.20  & 1339 \\
$\nu_5+\nu_8$& & ---  & 1363.34  & ---\\
$\nu_2+\nu_{12}$&  & --- & 1596.00 & 1600$^\dagger$ \\
$\nu_{13}$& e & 1721.09 & 1690.03 & 1692 \\
$\nu_3+\nu_{12}$& & --- & 1724.22 & 1728 \\
$\nu_5+\nu_{11}$&$\mathrm{f_1}$ & --- & 1753.22 & 1780 \\
$\nu_{14}$& $\mathrm{f_2}$& 1803.01 & 1769.45 & 1780 \\
$2\nu_{10}$& & ---  & 1967.22  & 1800-2200 \\
$\nu_{15}$& & 2333.77 & 2097.00 & 1800-2200 \\
$\nu_{10}+\nu_{11}$& & ---  & 2198.36 & 1800-2200 \\
\bottomrule
\end{tabular}
\label{sitab:oxd_assign}
\end{table}

\begin{table}[htbp]
\centering
\begin{tabular}{cccccc}
\toprule
\textbf{$N$} & \textbf{MP2 (sp)} & \textbf{CCSD(T) (sp)} & \multicolumn{3}{c}{\textbf{CCSD(T) (dp)}} \\
\cmidrule(lr){4-6}
 \textbf{beads} & \textbf{$\Delta_{\rm H}^{\rm sRPI}$ (cm$^{-1}$)} & $\Delta_{\rm H}^{\rm sRPI}$ \textbf{(cm$^{-1}$)} & $\Delta_{\rm H}^{\rm sRPI}$ \textbf{(cm$^{-1}$)} & $c_{\rm pc}$ & $\Delta_{\rm H}^{\rm pcRPI}$ \textbf{(cm$^{-1}$)}\\
\midrule
128 & &  & 26.12 & 0.93 & 24.36 \\
256 & &  & 32.33 & 1.02 & 32.84  \\
512 & 113.66 & 35.04 & 34.74 & 0.98 & 34.13  \\
1024 & 115.81 & 35.72 & 35.42 & 0.98 & 34.74  \\
2048 & 116.38 & 35.91 & 35.65 & 0.98* & 34.94*  \\
4096 & 116.51 & 35.95 & 35.69 & 0.98* & 34.98*  \\
\bottomrule

\end{tabular}
\caption{Numerical Convergence of the leading-order tunneling
  splittings. $N$ is the number of beads, and "sp" and "dp" refer to
  ML-PESs trained with single and double precision arithmetics,
  respectively. The entries labelled with * used the $c_{\rm pc}$
  value from the calculation with $N = 1024$.}
\label{sitab:convergence}
\end{table}

\end{document}